\documentclass[a4paper,11pt]{article}

\usepackage{jcappub} 

\usepackage[T1]{fontenc} 
\usepackage{comment}
\usepackage[toc,page]{appendix}

\let\originalleft\left
\let\originalright\right
\renewcommand{\left}{\mathopen{}\mathclose\bgroup\originalleft}
\renewcommand{\right}{\aftergroup\egroup\originalright}

\subheader{\hfill \normalsize YITP-25-76, IPMU25-0025, RIKEN-iTHEMS-Report-25}

\title{\boldmath Quasinormal Modes from EFT of Black Hole Perturbations in Vector-Tensor Gravity}

\author[a,*]{Shogo Tomizuka,\note[*]{Corresponding author.}}
\author[b]{Hajime Kobayashi,}
\author[b,c,d]{Naritaka Oshita,}
\author[e,b]{Kazufumi Takahashi,}
\author[b,f,g]{and Shinji Mukohyama}

\affiliation[a]{Department of physics, Kyoto University, 606-8502, Kyoto, Japan}
\affiliation[b]{Center for Gravitational Physics and Quantum Information, Yukawa Institute for Theoretical Physics, Kyoto University, 606-8502, Kyoto, Japan}
\affiliation[c]{The Hakubi Center for Advanced Research, Kyoto University, Yoshida Ushinomiyacho, Sakyo-ku, Kyoto 606-8501, Japan}
\affiliation[d]{RIKEN iTHEMS, Wako, Saitama, 351-0198, Japan}
\affiliation[e]{Department of Physics, College of Humanities and Sciences, Nihon University, Tokyo 156-8550, Japan}
\affiliation[f]{Research Center for the Early Universe (RESCEU), Graduate School of
Science, The University of Tokyo, Hongo 7-3-1, Bunkyo-ku, Tokyo
113-0033, Japan}
\affiliation[g]{Kavli Institute for the Physics and Mathematics of the Universe (WPI), The University of Tokyo Institutes for Advanced Study (UTIAS), The University of Tokyo, Kashiwa, Chiba 277-8583, Japan}

\abstract{We study the dynamics of odd-parity perturbations on a static and spherically symmetric black hole background with a timelike vector field based on the effective field theory (EFT) approach.
We derive the quadratic Lagrangian written in terms of two master variables, corresponding to the tensor and vector gravitons, which are coupled in general, while they can be decoupled on a stealth Schwarzschild(-de Sitter) background. For the stealth Schwarzschild background, we find that the quasinormal mode frequencies for both degrees of freedom are obtained from those in general relativity by simple scaling. Nonetheless, due to the fact that the metric perturbation is a non-trivial linear combination of the two degrees of freedom with different QNM spectra, the ringdown gravitational waves may exhibit characteristic modulation that can in principle be a signature of vector-tensor gravity.}

\begin{document}
\maketitle
\flushbottom

\section{Introduction}
\label{sec:intro}
General relativity (GR) is now commonly accepted as the standard model of gravity, and many attempts have been made to explain gravitational phenomena through its framework. 
However, as long as we rely on GR, there are phenomena that cannot be explained without assuming the existence of unknown entities such as dark matter and dark energy~\cite{SupernovaSearchTeam:1998fmf,SupernovaCosmologyProject:1998vns}.
These may instead suggest that gravity is not described by GR.
Scalar-tensor and vector-tensor theories are prominent alternatives to GR, introducing an additional field to modify gravitational interactions.
These theories diverge from GR by incorporating a scalar or vector field on top of the metric tensor, leading to distinct predictions for gravitational phenomena.
After the pioneering work~\cite{Horndeski:1974wa} of Horndeski, where the most general action of scalar-tensor theories with second-order Euler-Lagrange equations was found, the framework of scalar-tensor gravity has been extended to degenerate higher-order scalar-tensor (DHOST)~\cite{Langlois:2015cwa,Crisostomi:2016czh,BenAchour:2016fzp,Takahashi:2017pje,Langlois:2018jdg}, U-DHOST~\cite{DeFelice:2018ewo,DeFelice:2021hps,DeFelice:2022xvq}, and generalized disformal theories~\cite{Takahashi:2021ttd,Takahashi:2022mew,Takahashi:2023jro,Takahashi:2023vva}.
Vector-tensor gravity, by allowing the vector field to have a mass term, involves a longitudinal degree of freedom in addition to transverse modes of the vector field. 
It should be noted that such vector-tensor theories can be regarded as an extension of scalar-tensor theories.
Indeed, in the limit where the transverse modes are decoupled, one recovers shift-symmetric scalar-tensor theories.
Similarly to the case of scalar-tensor gravity, there have been extensive studies on generalizations of vector-tensor gravity, including generalized Proca theories~\cite{Tasinato:2014eka,Heisenberg:2014rta,Allys:2015sht,BeltranJimenez:2016rff,Allys:2016jaq} and others~\cite{Heisenberg:2016eld,Kimura:2016rzw,deRham:2020yet}.
In order to determine whether GR or alternative theories better describe our Universe, it is essential to observe strong-field phenomena such as gravitational waves from binary black hole mergers~\cite{LIGOScientific:2016aoc}.
Indeed, the gravitational waveform is affected by the underlying theory of gravity.
For instance, the tidal Love numbers and quasinormal mode frequencies (QNMFs)~\cite{Regge:1957td,Chandrasekhar:1975zza} depend on the underlying gravitational theory, and they would be encoded in the inspiral and ringdown phases, respectively.
Also, compared to current gravitational wave detectors, next-generation detectors such as LISA~\cite{LISA:2024hlh} are expected to improve the signal-to-noise ratio of gravitational wave observations, allowing for more detailed analyses of gravitational waves.
This improvement raises the possibility of detecting deviations from GR. 

The effective field theory (EFT) approach is useful for distinguishing between theories through observations, as it provides a universal description of perturbations about a given background.
In recent years, EFTs for scalar-tensor gravity with timelike scalar profile~\cite{Arkani-Hamed:2003pdi,Arkani-Hamed:2003juy,Cheung:2007st,Gubitosi:2012hu,Mukohyama:2022enj} and vector-tensor gravity with timelike vector profile~\cite{Cheng:2006us,Mukohyama:2006mm,Aoki:2021wew,Aoki:2023bmz} have been developed.
The key idea of the EFTs is that the additional scalar/vector field, when it has a non-trivial background, partially breaks the four-dimensional diffeomorphism invariance and defines a preferred slicing/threading of spacetime.
One can write down the most general action invariant under the transformations that preserve the preferred slicing/threading up to necessary orders in perturbations and derivatives, which can be used to describe the dynamics of the perturbations on the given background in a model-independent manner.
Among the EFTs developed so far, those in Refs.~\cite{Mukohyama:2022enj} and \cite{Aoki:2023bmz} apply to general background spacetimes with a timelike scalar field and a vector field, respectively, and can be used to study the perturbations of a black hole.
For scalar-tensor gravity, QNMFs~\cite{Mukohyama:2023xyf} and tidal Love numbers~\cite{Barura:2024uog} have been derived based on the EFT.
The aim of this paper is to calculate QNMFs using the EFT of vector-tensor gravity and investigate how they are modified by the vector field.

The rest of this paper is organized as follows.
In Section~\ref{sec:setup}, we give a brief review of the EFT of vector-tensor gravity on an arbitrary background with a timelike vector field~\cite{Aoki:2023bmz}.
In Section~\ref{sec:perturbation}, we derive the quadratic Lagrangian and two master variables describing the dynamics of odd-parity perturbations on a static and spherically symmetric background.
In Section~\ref{sec:QNM}, we calculate QNMFs for the stealth Schwarzschild solution where the two master variables are decoupled.
Finally, we draw our conclusions in Section~\ref{sec:conclusion}.

\section{Setup}\label{sec:setup}
\subsection{Review of the EFT}
In this subsection, we briefly review the construction of the EFT of perturbations on an arbitrary background with a timelike vector field~\cite{Aoki:2023bmz}. The essential idea of the EFT is that the background of a vector field~$v_\mu$ spontaneously breaks a part of the four-dimensional diffeomorphism invariance. 
In order to identify the symmetry breaking pattern of the EFT, it is convenient to introduce a St\"{u}ckelberg field~$\tilde{\tau}$ and rewrite the vector field as
\begin{align}
    v_\mu = \partial_\mu \tilde{\tau} +g_M A_\mu~,\quad
    g_M\equiv \frac{g}{M_*}~,
\end{align}
where $A_\mu$ is a gauge field, $g$ is a gauge coupling constant, and $M_*$ is some mass scale (e.g., the Planck mass).
The vector field~$v_\mu$ is invariant under the following $U(1)$ transformation:
\begin{align}
    \tilde{\tau}\rightarrow \tilde{\tau}+g_M\varphi~,\quad
    A_\mu\rightarrow A_\mu - \partial_\mu \varphi~,\label{eq:trans}
\end{align}
where $\varphi$ is an arbitrary scalar function. 
One can use this transformation to move to the unitary gauge where $\tilde{\tau}$ coincides with the time coordinate~$\tau$.
In this gauge, the vector field can be expressed as
\begin{align}
    v_\mu|_\mathrm{\tilde{\tau}=\tau}=\delta_\mu^\tau +g_MA_\mu\equiv\boldsymbol{\delta}^\tau_\mu~.
\end{align}
Here and in what follows, quantities in the unitary gauge which contain vector components through the gauge coupling~$g_M$ are represented in bold notation.
Note that, in this gauge, the transformation~(\ref{eq:trans}) can be seen as a combination of the time diffeomorphism and the $U(1)$ transformation, and the vector field is invariant under this combined transformation.
We call this transformation the residual $U(1)$ transformation.
Then, our EFT possesses the residual $U(1)$ symmetry and the spatial diffeomorphism invariance.
In what follows, we summarize the building blocks of the EFT action which respect these residual symmetries.

First, the norm of the vector field,
\begin{align}
    \boldsymbol{g}^{\tau\tau}\equiv g^{\mu\nu}\boldsymbol{\delta}_\mu^\tau\boldsymbol{\delta}_\nu^\tau = g^{\tau\tau}+2g_MA^\tau+g_M^2A_\mu A^\mu~,\label{eq:bold-g-tau-tau}
\end{align}
is invariant under the residual symmetries, and hence is a building block of the EFT.
Then, the unit (future-directed) timelike vector can be defined as
\begin{align}
    \boldsymbol{n}_\mu=-\frac{\boldsymbol{\delta}_\mu^\tau}{\sqrt{-\boldsymbol{g}^{\tau\tau}}}~,\label{eq:nmu}
\end{align}
where the norm of the vector field~$\boldsymbol{g}^{\tau\tau}<0$ is defined in Eq.~\eqref{eq:bold-g-tau-tau}.\footnote{Note that the direction defined by Eq.~(\ref{eq:nmu}) is different from the one defined by the time coordinate (i.e., $-\delta^\tau_\mu/\sqrt{-g^{\tau\tau}}$) because of the presence of the transverse component of $A_\mu$.
In particular, $\boldsymbol{n}_\mu$ is not hypersurface-orthogonal in general.
Therefore, it is not appropriate to use $3+1$ decomposition of spacetime based on the constant-$\tau$ slicing, and one has instead to use $1+3$ decomposition based on the preferred threading defined by $v_\mu$.
The details of the $1+3$ decomposition are presented in Appendix~A of~\cite{Aoki:2021wew}.}
The projection tensor can be defined by
\begin{align}
    \boldsymbol{h}_{\mu\nu}\equiv g_{\mu\nu}+\boldsymbol{n}_\mu\boldsymbol{n}_\nu~,
\end{align}
which satisfies $\boldsymbol{h}^\mu{}_{\alpha}\boldsymbol{h}^\alpha{}_{\nu}=\boldsymbol{h}^\mu{}_{\nu}$ and $\boldsymbol{h}^\alpha{}_{\mu}\boldsymbol{n}_\alpha=0$.
We can construct kinematical quantities associated with $\boldsymbol{n}_\mu$ as
\begin{align}
    \boldsymbol{K}_{\mu\nu}&\equiv\boldsymbol{h}_{(\mu|}{}^{\alpha}\nabla_\alpha\boldsymbol{n}_{|\nu)}~,\label{eq:Kmunu}\\
    \boldsymbol{\omega}_{\mu\nu}&\equiv \boldsymbol{h}_{[\mu|}{}^{\alpha}\nabla_\alpha \boldsymbol{n}_{|\nu]}~,\label{eq:omegamunu}\\
    \boldsymbol{a}_\mu&\equiv \boldsymbol{n}^\alpha\nabla_\alpha\boldsymbol{n}_\mu~,\label{eq:amu}
\end{align}
where the symbols~$(\,)$ and $[\,]$ respectively denote symmetrization and antisymmetrization of indices.
For the derivative operator, we can split it into components parallel and perpendicular to $\boldsymbol{n}_\mu$ as
\begin{align}
    \pounds_{\boldsymbol{n}}~,\quad
    \boldsymbol{\mathrm{D}}_\mu\equiv\boldsymbol{h}_\mu{}^{\alpha}\nabla_\alpha~,
\end{align}
where $\pounds_{\boldsymbol{n}}$ is the Lie derivative along $\boldsymbol{n}_\mu$.
Furthermore, using the projection tensor~$\boldsymbol{h}_{\mu\nu}$ and the covariant derivative~$\boldsymbol{\mathrm{D}}_\mu$, we can naturally define the orthogonal spatial curvature~${}^{(3)}\!\boldsymbol{R}_{\mu\nu\rho\sigma}$ on the projected space.
Also, from the Gauss equation, ${}^{(3)}\!\boldsymbol{R}_{\mu\nu\rho\sigma}$ 
is related to the four-dimensional curvature tensor~$\tilde{R}_{\mu\nu\rho\sigma}$ as follows:
\begin{align}
    \boldsymbol{h}_\mu{}^{\alpha}\boldsymbol{h}_\nu{}^{\beta}\boldsymbol{h}_\rho{}^{\gamma}\boldsymbol{h}_\sigma{}^{\delta}\tilde{R}_{\alpha\beta\gamma\delta}={}^{(3)}\!\boldsymbol{R}_{\mu\nu\rho\sigma}+2\boldsymbol{W}_{[\rho|\mu}\boldsymbol{W}_{|\sigma]\nu}~,
\end{align}
where $\boldsymbol{W}_{\mu\nu}$ is defined by 
\begin{align}
    \boldsymbol{W}_{\mu\nu}=\boldsymbol{K}_{\mu\nu}+\boldsymbol{\omega}_{\mu\nu}~.
\end{align}
We can also define the trace part of ${}^{(3)}\!\boldsymbol{R}_{\mu\nu\rho\sigma}$ as
\begin{align}
    {}^{(3)}\!\boldsymbol{R}_{\mu\nu}={}^{(3)}\!\boldsymbol{R}^\alpha{}_{(\mu|\alpha|\nu)}~.
\end{align}
Note that ${}^{(3)}\!\boldsymbol{R}_{\mu\nu\rho\sigma}$ defined here differs from the usual one associated with a hypersurface in that the antisymmetric part of the trace does not vanish in general.
However, this antisymmetric part can be expressed in terms of $\boldsymbol{W}_{\mu\nu}$, and therefore it does not need to be considered explicitly as a part of the building blocks of the EFT action.
It should also be noted that all the components of the four-dimensional curvature~$\tilde{R}_{\mu\nu\rho\sigma}$ can be written in terms of the kinematical quantities and the orthogonal spatial curvature, and hence is omitted from the independent building blocks.

Next, let us consider EFT building blocks constructed out of the gauge field~$A_\mu$.
From the field strength of the vector field~$F_{\mu\nu}=\partial_\mu A_\nu -\partial_\nu A_\mu$, which respects the residual symmetries, we define the electric and magnetic parts as follows:
\begin{align}
    \boldsymbol{E}_{\mu\nu}&\equiv \epsilon_{\mu\nu}{}^{\rho\sigma}\boldsymbol{n}_\rho F_{\sigma\alpha}\boldsymbol{n}^\alpha~,\\
    \boldsymbol{B}_{\mu\nu}&\equiv \boldsymbol{h}_\mu{}^{\alpha}\boldsymbol{h}_{\nu}{}^{\beta}F_{\alpha\beta}~.
\end{align}
Here, $\epsilon_{\mu\nu\rho\sigma}=\sqrt{-g}\,\varepsilon_{\mu\nu\rho\sigma}$ is the completely antisymmetric tensor, where $\varepsilon_{\mu\nu\rho\sigma}$ denotes the Levi-Civita symbol with $\varepsilon_{0123}=1$.
From Eqs.~(\ref{eq:omegamunu}) and (\ref{eq:amu}), we obtain the following relations among $\boldsymbol{\omega}_{\mu\nu}$, $\boldsymbol{a}_\mu$, $\boldsymbol{E}_{\mu\nu}$, and $\boldsymbol{B}_{\mu\nu}$:
\begin{align}
    \boldsymbol{\omega}_{\mu\nu}&=-\frac{g_M}{2\sqrt{-\boldsymbol{g}^{\tau\tau}}}\boldsymbol{B}_{\mu\nu}~,\\
    \boldsymbol{a}_\mu &=\frac{g_M}{2\sqrt{-\boldsymbol{g}^{\tau\tau}}}\epsilon_\mu{}^{\nu\alpha\beta}\boldsymbol{n}_\nu\boldsymbol{E}_{\alpha\beta}-\frac{\boldsymbol{h}_\mu{}^{\nu}\partial_\nu\boldsymbol{g}^{\tau\tau}}{2\boldsymbol{g}^{\tau\tau}}~.
\end{align}
Therefore, $\boldsymbol{\omega}_{\mu\nu}$ and $\boldsymbol{a}_\mu$ can be removed from the set of independent building blocks.

Using the building blocks described above, the EFT action under the unitary gauge can be expressed in terms of a general (scalar) function~$\mathcal{L}$ as follows:\footnote{Strictly speaking, one could include the Levi-Civita tensor as an additional building block in \eqref{eq:effectiveS}.
However, as clarified in \cite{Aoki:2023bmz}, a large class of vector-tensor theories (including a subclass of the generalized Proca theory) can be recast in the form of \eqref{eq:effectiveS}.}
\begin{align}
    S_{\rm EFT}=\int \mathrm{d}^4 x\sqrt{-g}\,\mathcal{L}(\boldsymbol{g}^{\tau\tau},{}^{(3)}\!\boldsymbol{R}^\mu{}_{\nu},\boldsymbol{K}^\mu{}_{\nu},\boldsymbol{E}^\mu{}_{\nu},\boldsymbol{B}^\mu{}_{\nu},\pounds_{\boldsymbol{n}},\boldsymbol{\mathrm{D}}_\mu)~.\label{eq:effectiveS}
\end{align}
Notice that the building blocks with two tensorial indices are included with one of the indices raised, so that one can simply use the Kronecker delta~$\delta^\mu_\nu$ to contract those indices.
Although the action~\eqref{eq:effectiveS} can be used to study the dynamics of perturbations on a given background in principle, it is useful to expand the action up to a necessary order in perturbations (and derivatives) in practice.
Here, we define perturbations of the building blocks as follows:
\begin{align}
    \delta\boldsymbol{g}^{\tau\tau}\equiv \boldsymbol{g}^{\tau\tau}-\bar{\boldsymbol{g}}^{\tau\tau}(\tau, \vec{x})~,\quad
    \delta\boldsymbol{Z}^{I\mu}{}_{\nu}\equiv\boldsymbol{Z}^{I\mu}{}_{\nu}-\bar{\boldsymbol{Z}}^{I\mu}{}_{\nu}(\tau,\vec{x})~,
\end{align}
where barred quantities denote background values, and $\boldsymbol{Z}^{I\mu}{}_{\nu}$ ($I=1,2,3,4$) collectively represents ${}^{(3)}\!\boldsymbol{R}^\mu{}_{\nu}$, $\boldsymbol{K}^\mu{}_{\nu}$, $\boldsymbol{E}^\mu{}_{\nu}$, and $\boldsymbol{B}^\mu{}_{\nu}$.
Note that, in order to accommodate inhomogeneous backgrounds such as black holes, we consider the background values of the EFT building blocks, which depend on both the time and spatial coordinates.
After some rearrangements, up to the second order in perturbations and the leading order in derivatives, the effective action~\eqref{eq:effectiveS} can be expressed as follows:
\begin{align}
 S_{\mathrm{EFT}} =  \int {\rm d}^4x \sqrt{-g} \bigg[&\frac{M^2_*}{2}f(x)\boldsymbol{R}-\Lambda(x)-c(x)\boldsymbol{g}^{\tau\tau}-d(x)\boldsymbol{K} -\alpha(x)^\mu{}_{\nu} \boldsymbol{K}^\nu{}_{\mu}-\beta(x)^\mu{}_{\nu} {}^{(3)}\!\boldsymbol{R}^\nu{}_{\mu} \nonumber \\   
 &-\xi_1(x)^\mu{}_{\nu}\boldsymbol{E}^\nu{}_{\mu}-\xi_2(x)^\mu{}_{\nu} \boldsymbol{B}^\nu{}_{\mu}  
 + \frac{1}{2}m_2^4(x)\left(\frac{\delta\boldsymbol{g}^{\tau\tau}}{-\bar{\boldsymbol{g}}^{\tau\tau}}\right)^2 \nonumber \\ 
 & +\lambda_2^I(x)^\mu{}_{\nu} \left(\frac{ \delta\boldsymbol{g}^{\tau\tau}}{-\bar{\boldsymbol{g}}^{\tau\tau}}\right)\delta \boldsymbol{Z}^{I\nu}{}_{\mu} + \frac{1}{2}\lambda_3^{IJ}(x)^{\mu\rho}{}_{\nu\sigma} \delta \boldsymbol{Z}^{I\nu}{}_{\mu}\delta\boldsymbol{Z}^{J\sigma}{}_{\rho}  \bigg]~,\label{eq:2ndaction}
\end{align}
where the summation over $I$ and $J$ is implicit, and we have omitted terms of higher order in perturbations and/or derivatives.
The coefficients of the EFT action are functions of $x=(\tau, \vec{x})$ in general,\footnote{Once the background is fixed, it is straightforward to see how each concrete theory is embedded in the EFT, and the relation between the parameters of the concrete theory and those of the EFT is often called ``dictionary''. A dictionary for generalized Proca theories was obtained in \cite{Aoki:2023bmz}.} among which those in front of tadpole terms (i.e., $f$, $\Lambda$, $c$, $d$, $\alpha^\mu{}_\nu$, $\beta^\mu{}_\nu$, $\xi_1{}^\mu{}_\nu$, and $\xi_2{}^\mu{}_\nu$) are chosen so that they are consistent with the tadpole cancellation conditions.
(See \cite{Aoki:2023bmz} for details.)
Note that, instead of the four-dimensional Ricci scalar~$\tilde{R}$, we have used $\boldsymbol{R}$ which denotes the Ricci scalar with the divergence term subtracted, i.e.,
\begin{align}
    \boldsymbol{R}\equiv {}^{(3)}\!\boldsymbol{R} + \boldsymbol{K}_{\mu\nu}\boldsymbol{K}^{\mu\nu}+\boldsymbol{\omega}^\mu{}_{\nu}\boldsymbol{\omega}^\nu{}_{\mu} - \boldsymbol{K}^2
    = \tilde{R} - 2\nabla_\mu(\boldsymbol{K} \boldsymbol{n}^\mu-\boldsymbol{n}^\nu\nabla_\nu \boldsymbol{n}^\mu)~.
\end{align}
It should also be noted that we choose $\alpha^\mu{}_{\nu}$ to be traceless, i.e., $\alpha^\mu{}_{\mu}=0$.

Finally, let us comment on the consistency relations that the EFT coefficients should satisfy.
Since each EFT coefficient is a function of $(\tau, \vec{x})$ in general, the action~\eqref{eq:2ndaction} apparently does not respect the residual symmetries of the EFT (i.e., the spatial diffeomorphism invariance and the residual $U(1)$ symmetry).
Nevertheless, the action does respect these symmetries (up to the second order in perturbations) as it is derived from Eq.~\eqref{eq:effectiveS} which manifestly respects the residual symmetries of the EFT.
Due to this fact, the EFT coefficients should satisfy two sets of consistency relations to ensure the spatial diffeomorphism invariance and the residual $U(1)$ symmetry of Eq.~\eqref{eq:2ndaction}, whose explicit form can be found in \cite{Aoki:2023bmz}.

\subsection{Static and spherically symmetric background}
Our main focus in this paper is perturbations about static and spherically symmetric backgrounds, described by the following metric:
\begin{align}
    \mathrm{d}s^2=-A(r)\mathrm{d}t^2+\frac{1}{B(r)}\mathrm{d}r^2+r^2 
    ({\rm d}\theta^2+\sin^2\theta\,{\rm d}\phi^2)~,\label{eq:metric}
\end{align}
where $A$ and $B$ are functions of $r$.
As we will see shortly, when considering perturbations on this background spacetime, it is convenient to use the Lema{\^i}tre (or Gaussian-normal) coordinates. In this coordinate system, the metric~(\ref{eq:metric}) is written as
\begin{align}
    \mathrm{d}s^2=-\mathrm{d}\tau^2 +[1-A(r)]\mathrm{d}\rho^2+
    r^2({\rm d}\theta^2+\sin^2\theta\,{\rm d}\phi^2)~.\label{eq:lemaitremetric}
\end{align}
Here, the original coordinates~$\{t,r\}$ and the new coordinates~$\{\tau,\rho\}$ are related to each other via
\begin{align}
    \mathrm{d}\tau=\mathrm{d}t+\sqrt{\frac{1-A}{AB}}\,\mathrm{d}r~,\quad
    \mathrm{d}\rho=\mathrm{d}t+\frac{\mathrm{d}r}{\sqrt{AB(1-A)}}~.
\end{align}
Note that the areal radius~$r$ is a function of $\tau$ and $\rho$.
In particular, we have
\begin{align}
    \partial_\rho r=-\dot{r}=\sqrt{\frac{B(1-A)}{A}}~,
\end{align}
with a dot denoting the derivative with respect to $\tau$, meaning that $r$ is a function of $\rho-\tau$.

In what follows, for simplicity, we consider static and spherically symmetric backgrounds with
\begin{align}
    \bar{X} \equiv -\frac{1}{2}\bar{A}_\mu \bar{A}^\mu = \frac{q^2}{2}~,\label{eq:veccond}
\end{align}
where $q$ is a constant. For the background metric~\eqref{eq:lemaitremetric}, one possible solution satisfying this condition is given by
\begin{align}
    \bar{A}_\tau=q~, \quad
\bar{A}_\rho =0~.\label{eq:vecback}
\end{align}
Note that a nontrivial $r$-dependence of $\bar{X}$ is not prohibited by the symmetry.
Nevertheless, we impose the condition~\eqref{eq:veccond} as such a configuration has been studied in \cite{Cheng:2006us,Chagoya:2016aar,Minamitsuji:2017aan,Minamitsuji:2021gcq,Atkins:2023axs} in relation to the stealth solutions (i.e., those with the same metric as in GR) in scalar-tensor theories~\cite{Mukohyama:2005rw,Babichev:2013cya,Kobayashi:2014eva,Motohashi:2018wdq,BenAchour:2018dap,Motohashi:2019sen,Takahashi:2020hso}.
Furthermore, we consider the EFT action~\eqref{eq:2ndaction} with all the EFT parameters being functions only of $r$, in order to be consistent with the static and spherically symmetric background.
Also, we assume $f=1$, $\beta^\mu{}_\nu=\xi_1{}^\mu{}_\nu=\xi_2{}^\mu{}_\nu=0$, and $\alpha^\mu{}_{\nu} = \alpha(r) (\bar{\boldsymbol K}^\mu{}_{\nu}-\bar{\boldsymbol{K}}\delta^\mu{}_{\nu}/4)$ with $\alpha$ being a function of $r$.\footnote{One can set $f=1$ without loss of generality, as any non-trivial $f$ can be absorbed via a conformal transformation of the metric.
Also, we adopt the restrictions on the EFT parameters that are satisfied by (a subclass of) the generalized Proca theory, as derived in \cite{Aoki:2023bmz}.
These restrictions imply that $\beta^\mu{}_\nu=0$, $\alpha^\mu{}_{\nu}\propto \bar{\boldsymbol K}^\mu{}_{\nu}-\bar{\boldsymbol{K}}\delta^\mu{}_{\nu}/4$, $\xi_1{}^\mu{}_\nu\propto \bar{\boldsymbol E}^\mu{}_{\nu}$, and $\xi_2{}^\mu{}_\nu\propto \bar{\boldsymbol B}^\mu{}_{\nu}$.
Since $\bar{\boldsymbol E}^\mu{}_{\nu}$ and $\bar{\boldsymbol B}^\mu{}_{\nu}$ vanish on the background of our interest, it follows that $\xi_1{}^\mu{}_\nu=\xi_2{}^\mu{}_\nu=0$.}
With these assumptions, our EFT action takes the following form:
\begin{align}
    S_{(2)}&=\int \mathrm{d}^4x\sqrt{-g}\bigg[\frac{M_*^2}{2}\boldsymbol{R}-\Lambda (r)-c(r)\boldsymbol{g}^{\tau\tau}-\Tilde{d}(r)\boldsymbol{K}-\alpha (r)\Bar{\boldsymbol{K}}^\mu_\nu\boldsymbol{K}^\nu_\mu \nonumber \\
    &\quad +\frac{1}{2}m_2^4(r)\left(\frac{\delta\boldsymbol{g}^{\tau\tau}}{-\bar{\boldsymbol{g}}^{\tau\tau}}\right)^2+\lambda_2^I(r)^\mu{}_{\nu} \left(\frac{ \delta\boldsymbol{g}^{\tau\tau}}{-\bar{\boldsymbol{g}}^{\tau\tau}}\right)\delta \boldsymbol{Z}^{I\nu}{}_{\mu} + \frac{1}{2}\lambda_3^{IJ}(r)^{\mu\rho}{}_{\nu\sigma} \delta \boldsymbol{Z}^{I\nu}{}_{\mu}\delta\boldsymbol{Z}^{J\sigma}{}_{\rho} \bigg] ~,
    \label{eq:2ndactionspherical}
\end{align}
where we have defined $\tilde{d}(r)\equiv d(r)-\alpha(r)\bar{\boldsymbol{K}}/4$.
Note that since the background vector field~(\ref{eq:vecback}) is constant, the background quantities~$\bar{\boldsymbol{E}}_{\mu\nu}$ and $\bar{\boldsymbol{B}}_{\mu\nu}$ are vanishing. 
From the variation of the action~\eqref{eq:2ndactionspherical} with respect to the metric, we obtain the tadpole cancellation condition (or the Einstein equation) as
\begin{align}
    M^2_* \bar{G}_{\mu\nu} = \bar{T}_{\mu\nu}~,\label{eq:ein}
\end{align}
where the effective energy-momentum tensor~$\bar{T}_{\mu\nu}$ is given by
\begin{align}
    \bar{T}_{\mu\nu} = & -(\Lambda + c \bar{\boldsymbol g}^{\tau\tau} - \bar{\boldsymbol n}^\lambda \partial_\lambda \tilde{d} + \alpha \bar{\boldsymbol K}^\lambda{}_{\sigma}\bar{\boldsymbol K}^{\sigma}{}_{\lambda} ) \bar{g}_{\mu\nu} -(2c\bar{\boldsymbol{g}}^{\tau\tau}+\bar{\boldsymbol n}^\lambda \partial_\lambda \tilde{d} - \alpha \bar{\boldsymbol K}^\lambda{}_{\sigma}\bar{\boldsymbol K}^{\sigma}{}_{\lambda})\bar{\boldsymbol{n}}_\mu\bar{\boldsymbol{n}}_\nu \nonumber\\&
    - 2  \bar{\boldsymbol n}_{(\mu}\partial_{\nu)} \tilde{d}+ 2 \alpha \bar{\boldsymbol K}^\sigma{}_{\mu} \bar{\boldsymbol K}_{\sigma \nu}
    - 2\bar{\nabla}_\lambda (\alpha \bar{\boldsymbol K}^\lambda{}_{(\mu} \bar{\boldsymbol n}_{\nu)}) + \bar{\nabla}_\lambda (\alpha \bar{\boldsymbol K}_{\mu\nu} \bar{\boldsymbol n}^\lambda)~. \label{Tmn}
\end{align}
Note that the non-vanishing components of the background Einstein tensor~$\bar{G}^{\mu}{}_{\nu}$ are
\begin{align}
    \begin{split}
    \bar{G}^\tau{}_\tau&=-\frac{[r(1-B)]'}{r^2}+\frac{1-A}{r}\left(\frac{B}{A}\right)'~, \\
    \bar{G}^\tau{}_\rho&=-\frac{1-A}{r}\left(\frac{B}{A}\right)'~, \\
    \bar{G}^\rho{}_\rho&=-\frac{[r(1-B)]'}{r^2}-\frac{1}{r}\left(\frac{B}{A}\right)'~, \\
    \bar{G}^\theta{}_\theta&=\frac{B(r^2A')'}{2r^2A}+\frac{(r^2A)'}{4r^2}\left(\frac{B}{A}\right)'~,
    \end{split}
\end{align}
with a prime denoting the derivative with respect to $r$.
Also, it should be noted that we have $\bar{\boldsymbol{n}}_\mu=-\delta^\tau_\mu/\sqrt{-\bar{g}^{\tau\tau}}$ on the background of our interest, and therefore $\bar{\boldsymbol{n}}_\mu$ is normal to a constant-$\tau$ hypersurface.
In particular, this implies that $\bar{\boldsymbol{K}}_{\mu\nu}$ in Eq.~\eqref{Tmn} can be replaced by the extrinsic curvature on a constant-$\tau$ hypersurface.
Then, independent components of Eq.~\eqref{eq:ein} can be written as
\begin{align}
\Lambda- (1 + g_M q)^2c	&=M_*^2(\bar{G}^\tau{}_\rho-\bar{G}^\rho{}_\rho)~, \label{eq:ein1}\\
\Lambda + (1 + g_M q)^2c &=-M_*^2\bar{G}^\tau{}_\tau~,\label{eq:ein2} \\			\left[\partial_\rho\bar{K}+\frac{1-A}{r}\left(\frac{B}{A}\right)'\,\right]\alpha+\frac{A'B}{2A}\alpha'+\sqrt{\frac{B(1 - A)}{A}}\, \tilde{d}^{\,\prime}&=-M_*^2\bar{G}^\tau{}_\rho~,\label{eq:ein3} \\
\frac{1}{2r^2}\sqrt{\frac{B}{A}}\left[r^4\sqrt{\frac{B}{A}}\left(\frac{1-A}{r^2}\right)'\alpha\right]'	&=M_*^2(\bar{G}^\rho{}_\rho-\bar{G}^\theta{}_\theta)~,\label{eq:ein4}
\end{align}
where $\bar{K}$ is the background value of the trace of the extrinsic curvature on a constant-$\tau$ hypersurface, i.e.,
\begin{align}
    \bar{K}=-\frac{2}{r}\sqrt{\frac{B}{A(1-A)}}\bigg(1-A-\frac{rA'}{4}\bigg)~.
\end{align}
Meanwhile, the variation with respect to $A_\tau$ (i.e., the temporal component of the gauge field) yields
\begin{align}
    g_M (1 + g_Mq) c= 0~,\label{eq:veceom}
\end{align}
which implies $c=0$ so long as $g_M(1+g_M q)\neq 0$.
Combining this with Eqs.~\eqref{eq:ein1} and \eqref{eq:ein2}, we find $A\propto B$, and hence we can set $A=B$ by an appropriate rescaling of the time coordinate.
Then, Eq.~\eqref{eq:ein4} can be integrated to yield
\begin{align}\label{alpha}
    \alpha = M_*^2 + \frac{3\lambda}{r(2-2A+rA^\prime)}~,
\end{align}
with $\lambda$ being an integration constant.

Having derived the tadpole cancellation conditions, we are now ready to investigate perturbations on the background based on the EFT.
Note in passing that we have confirmed the compatibility of the above tadpole cancellation conditions and the consistency relations, where the latter ensure that the EFT respects the residual symmetries (see \cite{Aoki:2023bmz} for details).

\section{Odd-parity perturbations}\label{sec:perturbation}
In this section, we consider the dynamics of odd-parity perturbations based on the EFT of vector-tensor gravity.
First, we introduce odd-parity perturbations for the metric and the gauge field.
The odd-parity part of metric perturbation~$\delta g_{\mu\nu}\equiv g_{\mu\nu}-\bar{g}_{\mu\nu}$ can be expressed as
\begin{align}
\begin{split}
&\delta g_{\tau\tau}=\delta g_{\tau\rho} 
=\delta g_{\rho\rho}=0~,\\
&\delta g_{\tau a}
=\sum_{l,m}r^2 h_{0,lm}(\tau,\rho)E_a{}^{b}\bar{\nabla}_b Y_{lm}(\theta,\phi)~,\\
&\delta g_{\rho a}
=\sum_{l,m}r^2h_{1,lm}(\tau,\rho)E_a{}^{b}\bar{\nabla}_bY_{lm}(\theta,\phi)~,\\
&\delta g_{ab}=\sum_{l,m}r^2 h_{2,lm}(\tau,\rho)E_{(a|}{}^{c}\bar{\nabla}_c\bar{\nabla}_{|b)}Y_{lm}(\theta,\phi)~,
\end{split} \label{eq:metricpert}
\end{align}
where $Y_{lm}$ is the spherical harmonics.
Here, the indices~$a$, $b$, and $c$ denote $\{\theta,\phi\}$,  $E_{ab}$ is the antisymmetric tensor defined on a two-sphere, and $\bar{\nabla}_a$ is the covariant derivative on the sphere. 
Similarly, the odd-parity part of the vector field perturbation~$\delta A_\mu \equiv A_\mu - \bar{A}_\mu$ is
\begin{align}
\begin{split}
    &\delta A_\tau =\delta A_\rho =0~,\\
    &\delta A_{ a}=\sum_{l,m}r^2 a_{lm}(\tau,\rho)E_a{}^{b}\bar{\nabla}_b Y_{lm}(\theta,\phi)~.
\end{split}\label{eq:vecpert}
\end{align}
Thanks to the spherical symmetry of the background, modes with different $(l,m)$ evolve independently, and therefore we shall omit the indices~$l$ and $m$ in what follows.
Moreover, one can fix $m=0$ without loss of generality, and hence we expand the perturbations in terms of the Legendre polynomials~$P_l(\cos\theta)$ in practical computations.

One can set one of the four perturbation variables (i.e., $h_0$, $h_1$, $h_2$, and $a$) to zero by using the gauge degree of freedom of the odd-parity sector.
(See, e.g., \cite{Mukohyama:2022skk} for details.)
For $l\ge 2$, we set $h_2=0$, which is a complete gauge fixing and can be imposed at the level of Lagrangian~\cite{Motohashi:2016prk}.
For $l=1$, the variable $h_2$ is intrinsically absent, and we will set $h_1=0$.
Note, however, that this is an incomplete gauge fixing, and therefore we will impose it after deriving the equations of motion.

\subsection{\texorpdfstring{Odd-parity perturbations with $l\ge2$}{Odd-parity perturbations with l=2 or larger}}\label{ssec:odd}
In this subsection, we derive master equations for the odd-parity perturbations for $l\ge 2$ based on the EFT action~\eqref{eq:2ndactionspherical}, following~\cite{Mukohyama:2022skk,Mukohyama:2023xyf}. 
It should be noted that parity-even operators such as $\delta\boldsymbol{g}^{\tau\tau}$ and $\delta\boldsymbol{K}$ are at least of second order in odd-parity perturbations, and we can safely omit quadratic operators constructed out of them (e.g., $\delta\boldsymbol{g}^{\tau\tau}\delta\boldsymbol{K}$) in the linear perturbation analysis of the odd-parity sector.
Furthermore, in the present paper, we omit parity-violating operators such as $\delta\boldsymbol{E}^\mu{}_{\nu}\delta\boldsymbol{B}^\nu{}_{\mu}$ for simplicity.
Then, the relevant EFT action for the linear odd-parity perturbations is given by
\begin{align}
    S_{\rm odd}=\int\mathrm{d}^4x\sqrt{-g}\bigg[&\frac{M_*^2}{2}\boldsymbol{R}-\Lambda (r)-c(r)\boldsymbol{g}^{\tau\tau}-\tilde{d}(r)\boldsymbol{K}-\alpha (r)\bar{\boldsymbol{K}}^\mu_\nu\boldsymbol{K}^\nu_\mu \nonumber \\
    &+\frac{M_3^2(r)}{2}\delta\boldsymbol{K}^\mu_\nu\delta\boldsymbol{K}^\nu_\mu +\frac{\gamma_1(r)}{4}\delta\boldsymbol{E}^\mu{}_{\nu}\delta\boldsymbol{E}^\nu{}_{\mu}+\frac{\gamma_2(r)}{4}\delta\boldsymbol{B}^\mu{}_{\nu}\delta\boldsymbol{B}^\nu{}_{\mu}\bigg]~. \label{eq:oddeffectiveaction}
\end{align}
Note that the terms in the second line arise from the last term of Eq.~\eqref{eq:2ndactionspherical}, and we have introduced the ($r$-dependent) coefficients~$M_3^2$, $\gamma_1$, and $\gamma_2$ for notational simplicity.
One can then straightforwardly obtain the quadratic Lagrangian by plugging the backgrounds~\eqref{eq:lemaitremetric} and \eqref{eq:vecback} as well as perturbations~\eqref{eq:metricpert} and \eqref{eq:vecpert} into Eq.~\eqref{eq:oddeffectiveaction} and expanding it up to the second order in perturbations.
After performing the integration over the angular variables, we arrive at the following quadratic Lagrangian for $h_0$, $h_1$, and $a$:
\begin{align}
    \frac{2l+1}{2\pi j^2}\mathcal{L}_2&=p_1h_0^2+p_2h_1^2+p_3(\dot{h}_{1}-\partial_\rho h_0-p_4h_1+p_5\partial_\rho a)^2+p_6a^2+p_7ah_0 \nonumber \\
    &\quad+p_8h_1\partial_\rho a+p_9\dot{a}^2+p_{10}(\partial_\rho a)^2~, \label{qlag1}
\end{align}
up to total derivatives, where we have defined $j^2\equiv l(l+1)$.
The coefficients~$p_i$ are functions of $r$, whose explicit expression can be found in the \hyperref[app]{Appendix}.
We then introduce an auxiliary field~$\chi$ as
\begin{align}
    \frac{2l+1}{2\pi j^2}\mathcal{L}_2&=p_1h_0^2+p_2h_1^2+p_3\left[-\chi^2+2\chi(\dot{h}_{1}-\partial_\rho h_0-p_4h_1+p_5\partial_\rho a)\right]+p_6a^2+p_7ah_0 \nonumber \\
    &\quad+p_8h_1\partial_\rho a+p_9\dot{a}^2+p_{10}(\partial_\rho a)^2~. \label{qlag2}
\end{align}
Notice that the equation of motion for $\chi$ yields $\chi=\dot{h}_{1}-\partial_\rho h_0-p_4h_1+p_5\partial_\rho a$, which can be substituted back into \eqref{qlag2} to recover the original quadratic Lagrangian~\eqref{qlag1}.
From \eqref{qlag2}, one can derive the equations of motion for $h_0$ and $h_1$, which yield
\begin{align}
\begin{split}
    h_0&=-\frac{2\partial_\rho (p_3\chi)+p_7 a}{2p_1}~,\\
    h_1&=\frac{2(p_3\chi)^{\boldsymbol{\cdot}}+2p_3p_4\chi-p_8 \partial_\rho a}{2p_2}~.
\end{split} \label{h0h1}
\end{align}
Plugging these into \eqref{qlag2}, we arrive at the following Lagrangian written in terms of $\chi$ and $a$:
\begin{align}
    \frac{(j^2-2)(2l+1)}{2\pi j^2}\mathcal{L}_2&=s_1\dot{\chi}^2-s_2(\partial_\rho\chi)^2-s_3\chi^2+s_4\dot{a}^2-s_5(\partial_\rho a)^2-s_6a^2 \nonumber \\
    &\quad +s_7\dot{\chi}\partial_\rho a+s_8\chi\partial_\rho a+s_9\chi a~.\label{eq:lag4}
\end{align} 
The expression of the coefficients~$s_i$ can be found in the \hyperref[app]{Appendix}.

Let us comment on the coefficient~$p_4$ in Eq.~\eqref{qlag1} in relation to the EFT of black hole perturbations with timelike scalar profile~\cite{Mukohyama:2022enj,Mukohyama:2022skk}.
In the case of vanishing gauge coupling~$g_M$, one can confirm that the vector perturbation~$a$ is completely decoupled from the metric perturbations~$h_0$ and $h_1$.
Then, the metric sector reduces to nothing but the quadratic Lagrangian for odd modes in the EFT of scalar-tensor theories~\cite{Mukohyama:2022skk}.
(Note that the coefficients~$p_5$, $p_7$, $p_8$ of the mixing terms vanish when $g_M=0$.)
In \cite{Mukohyama:2022skk}, it was shown that a non-vanishing $p_4$ leads to the absence of slowly rotating black holes (or the divergence of the radial sound speed at spatial infinity), and therefore the condition~$p_4=0$ was imposed.
Similarly, in the EFT of vector-tensor theories with non-vanishing gauge coupling, we will show in Subsection~\ref{sec:dipole} that dipole perturbations are ill-behaved for $p_4\ne 0$ in general.
For this reason, we impose $p_4=0$ in what follows.

Moreover, it should be noted that the two degrees of freedom in the odd-parity sector, represented by $\chi$ and $a$, are coupled with each other in general, which would lead to interesting phenomena.\footnote{For example, it was pointed out in \cite{Cardoso:2024qie} that energy extraction from a non-spinning black hole and a characteristic late-time relaxation occur when multiple fields with different horizon radii are coupled with each other.}
Having said that, in what follows, we will restrict ourselves to the case where $\chi$ and $a$ are decoupled for simplicity.
Once the condition~$p_4=0$ (or, equivalently, $M_3^2=-\alpha$) is imposed, the only non-vanishing mixing term in Eq.~\eqref{eq:lag4} is the term with $s_9$, which can be written in terms of the EFT parameters as
    \begin{align}
    s_9=-\frac{(j^2-2)g_M M_*^2 r^4 \alpha'}{(1+g_M q)^2(M_*^2-\alpha)}~.
    \end{align}
Therefore, we impose $\alpha={\rm const}$ so that the mixing term vanishes.\footnote{Note in passing that this is the case with generalized Proca theories~\cite{Aoki:2023bmz}.
The EFT parameter~$\alpha$ is related to the non-minimal coupling with the Ricci scalar (i.e., the non-trivial $G_4$ function in \cite{Heisenberg:2014rta}) in generalized Proca theories.}
In this case, Eq.~\eqref{alpha} implies that the metric function~$A(r)$ should be of the Schwarzschild(-de Sitter) form.

Let us study the conditions for the absence of ghost and gradient instabilities under $p_4=0$ and $\alpha=-M_3^2={\rm const}$, where the dynamics of $\chi$ and $a$ is governed by the following Lagrangian:\begin{align}
    \frac{(j^2-2)(2l+1)}{2\pi j^2}\mathcal{L}_2=s_1\dot{\chi}^2-s_2(\partial_\rho\chi)^2-s_3\chi^2+s_4\dot{a}^2-s_5(\partial_\rho a)^2-s_6a^2~.\label{eq:lag4'}
\end{align}
For this purpose, we compute the squared sound speeds of $\chi$ and $a$.
The squared sound speeds in the $\rho$-direction are given by
    \begin{align}
    \begin{split}
    c_{\rho,\chi}^2&=\frac{\bar{g}_{\rho\rho}}{|\bar{g}_{\tau\tau}|}\frac{s_2}{s_1}
    =\frac{M_*^2}{M_*^2-\alpha}~, \\
    c_{\rho,a}^2&=\frac{\bar{g}_{\rho\rho}}{|\bar{g}_{\tau\tau}|}\frac{s_5}{s_4}
    =-\frac{g_M^2M_*^2\alpha}{2\gamma_1 (1+g_Mq)^2(M_*^2-\alpha)}-\frac{\gamma_2}{\gamma_1}~.
    \end{split}
    \end{align}
Similarly, the squared sound speeds in the $\theta$-direction are given by
    \begin{align}
    \begin{split}
    c_{\theta,\chi}^2&=\lim_{l\to\infty}\frac{r^2}{|\bar{g}_{\tau\tau}|}\frac{s_3}{j^2s_1}
    =\frac{M_*^2}{M_*^2-\alpha}~, \\
    c_{\theta,a}^2&=\lim_{l\to\infty}\frac{r^2}{|\bar{g}_{\tau\tau}|}\frac{s_6}{j^2s_4}
    =-\frac{g_M^2M_*^2\alpha}{2\gamma_1 (1+g_Mq)^2(M_*^2-\alpha)}-\frac{\gamma_2}{\gamma_1}~.
    \end{split}
    \end{align}
Interestingly, we have $c_\rho^2=c_\theta^2$ for both $\chi$ and $a$.
Therefore, we define $c_T^2\equiv c_{\rho,\chi}^2=c_{\theta,\chi}^2$ and $c_V^2\equiv c_{\rho,a}^2=c_{\theta,a}^2$, where the subscripts~$T$ and $V$ are meant to imply tensor and vector gravitons, respectively.
Indeed, in the decoupling limit~$g_M\to 0$, the variable~$\chi$ reduces to the master variable of the standard Regge-Wheeler equation in GR.
Also, for later convenience, we define the deviation of the squared sound speeds from unity as follows:\footnote{For the deviation of $c_V^2$ from unity, we use the notation~$\beta_V$ rather than $\alpha_V$, as the latter has been used in the EFT of vector-tensor theories on a cosmological background~\cite{Aoki:2021wew,Aoki:2024ktc} to parameterize the vectorial nature of dark energy.}
\begin{align}
    \alpha_T&\equiv c_T^2-1=\frac{\alpha}{M_*^2-\alpha}~, \label{eq:alpha}\\
    \beta_V&\equiv c_V^2 -1 =-\frac{g_M^2M_*^2\alpha_T}{2\gamma_1 (1+g_Mq)^2}-\frac{\gamma_1+\gamma_2}{\gamma_1}~. \label{betaV}
\end{align}
We are now ready to write down the stability conditions as follows:
    \begin{align}
    s_1>0~, \quad
    s_4>0~, \quad
    c_T^2>0~, \quad
    c_V^2>0~. \quad
    \end{align}
These conditions can be expressed in terms of the EFT coefficients as
\begin{align}
    M_*^2+M_3^2>0~, \quad
    M_*^2>0~, \quad
    \gamma_1>0~, \quad
    \gamma_2<-\frac{g_M^2\alpha M_*^2}{2(1+g_Mq)^2(M_*^2-\alpha)}~.
    \label{eq:stability}
\end{align}
Recall that one can prepare a dictionary for each concrete vector-tensor theory which relates the parameters of the theory to the EFT parameters.
By use of the dictionary, one can translate the stability conditions above in the language of the concrete theory.

Before proceeding further, let us comment on the metric reconstruction.
Although we refer to it as the tensor graviton, it involves the vector perturbation~$a$ in its definition in general.
As a result, the original metric perturbation variables~$h_0$ and $h_1$ are non-trivial linear combinations of $\chi$, $a$, and their derivatives [see Eq.~\eqref{h0h1}].
It should be noted that the tensor graviton~$\chi$ and the vector graviton~$a$ have different QNM spectra, and therefore a non-trivial modulation or the beating phenomenon may appear in the ringdown gravitational waves \cite{Cardoso:2024qie} (see Figure~\ref{fig:QNM_beating_Schwarzschild}).

Next, in order to obtain master equations in the standard form of wave equations, we express the quadratic Lagrangian~(\ref{eq:lag4'}) in terms of the Schwarzschild coordinates~$t$ and $r$. 
Note that one has to take into account the Jacobian determinant of the coordinate transformation to define the Lagrangian in the Schwarzschild coordinate system as
    \begin{align}
    \tilde{\cal L}_2\equiv \left|\frac{\partial(\tau,\rho)}{\partial(t,r)}\right|{\cal L}_2=\frac{{\cal L}_2}{\sqrt{1-A}}~,
    \end{align}
where we recall that $A=B$ is assumed.
Written explicitly, we have
\begin{align}
    \frac{(j^2-2)(2l+1)}{2\pi j^2}\tilde{\mathcal{L}_2} &= a_1(\partial_t\chi)^2-a_2(\partial_r\chi)^2+2a_3(\partial_t\chi)(\partial
    _r\chi)-a_4\chi^2 \nonumber \\
    &\quad +a_5(\partial_t a)^2-a_6(\partial_r a)^2+2a_7 (\partial_t a)(\partial_r a)-a_8a^2~, \label{eq:lag5}
\end{align}
where the coefficients~$a_i$ are given by
    \begin{align}
    \begin{split}
    a_1=\frac{s_1-(1-A)^2s_2}{A^2\sqrt{1-A}}~, \quad
    a_2=\sqrt{1-A}(s_2-s_1)~, \quad
    a_3=\frac{(1-A)s_2-s_1}{A}~, \quad
    a_4=\frac{s_3}{\sqrt{1-A}}~, \\
    a_5=\frac{s_4-(1-A)^2s_5}{A^2\sqrt{1-A}}~, \quad
    a_6=\sqrt{1-A}(s_5-s_4)~, \quad
    a_7=\frac{(1-A)s_5-s_4}{A}~, \quad
    a_8=\frac{s_6}{\sqrt{1-A}}~.
    \end{split}
    \end{align}
We shall derive the master equations for the tensor/vector gravitons in separate subsections.

\subsubsection{Master equation for tensor graviton}\label{sssec:tensor}
First, we consider the master equation for the tensor graviton~$\chi$, following~\cite{Mukohyama:2022enj,Mukohyama:2023xyf}.
In order to remove the cross-term~$(\partial_t \chi)(\partial_r \chi)$ from the $\chi$-sector of the quadratic Lagrangian~\eqref{eq:lag5}, we introduce the following time coordinate:
\begin{align}
    \tilde{t}_T\equiv t+\int \frac{a_3}{a_2}\mathrm{d}r=t+\int 
    \frac{\sqrt{1-A}}{A}\frac{\alpha_T}{A+\alpha_T}\mathrm{d}r~.
\end{align}
Using this time coordinate, the quadratic Lagrangian for the tensor graviton can be written as
\begin{align}
     \frac{(j^2-2)(2l+1)}{2\pi j^2}\tilde{\mathcal{L}}_{2,T}=\tilde{a}_1(\partial_{\tilde{t}_T}\chi)^2-a_2(\partial_r \chi)^2 -a_4\chi^2~,\label{eq:tenlag2}
\end{align}
where we have defined $\tilde{a}_1\equiv a_1+a_3^2/a_2$.
The tortoise coordinate for the tensor graviton is defined by
\begin{align}
    r_{*,T}\equiv\int\frac{{\rm d}r}{F_T(r)}~, \quad
    F_T(r)\equiv \sqrt{\frac{a_2}{\tilde{a}_1}}
    =\frac{A+\alpha_T}{\sqrt{1+\alpha_T}}~.
    \label{eq:tensorrs}
\end{align}
Note that the horizon radius for the effective metric of the tensor graviton, denoted by $r=r_T$, satisfies $F_T(r_T)=0$.
In terms of a new variable,
\begin{align}
    \Psi_T \equiv (\tilde{a}_1a_2)^{1/4}\chi~,
\end{align}
the master equation can be written in the form of a wave equation, i.e.,
\begin{align}
    \frac{\partial^2 \Psi_T}{\partial r_{*,T}^2}-\frac{\partial^2 \Psi_T}{\partial \tilde{t}_T^2}-V_{\mathrm{eff},T}(r)\Psi_T=0~,\label{eq:tenmas1}
\end{align}
where the effective potential for the tensor graviton is given by
\begin{align}
    V_{\mathrm{eff},T}&=\frac{a_4}{\tilde{a}_1}+\frac{1}{2\sqrt{\tilde{a}_1a_2}}\frac{{\rm d}^2\sqrt{\tilde{a}_1a_2}}{{\rm d}r_{*,T}^2}-\frac{1}{4\tilde{a}_1a_2}\left(\frac{{\rm d}\sqrt{\tilde{a}_1a_2}}{{\rm d}r_{*,T}}\right)^2\nonumber\\
    &=\sqrt{1+\alpha_T}\,F_T\left\{\frac{l(l+1)-2}{r^2}+\frac{r}{(1+\alpha_T)^{3/4}}\left[F_T\left(\frac{(1+\alpha_T)^{1/4}}{r}\right)'\,\right]'\,\right\}~.
    \label{eq:tenpot1}
\end{align}
Recall that a prime denotes the derivative with respect to $r$.
In the frequency domain, the master equation reduces to
\begin{align}
    \frac{{\rm d}^2Q_T}{{\rm d}r_{*,T}^2}
    +(\omega^2-V_{\mathrm{eff},T})
    Q_T=0~,\label{eq:tensormastereqfreq}
\end{align}
where we have denoted the master variable in the frequency domain as $Q_T$.
The new time coordinate $\tilde{t}_T$ differs from the Killing time $t$ only by a function of $r$.
Therefore, the frequency conjugate to $\tilde{t}_T$ coincides with that conjugate to $t$, which is the natural time coordinate for a static observer at spatial infinity.
When $\alpha_T=0$, the effective potential~(\ref{eq:tenpot1}) reduces to the conventional Regge-Wheeler potential for odd-parity metric perturbations in GR.

\subsubsection{Master equation for vector graviton}\label{sssec:vector}
Next, we consider the master equation for the vector graviton~$a$.
In order to remove the cross-term~$(\partial_t a)(\partial_r a)$ from the $a$-sector of the quadratic Lagrangian~\eqref{eq:lag5}, we introduce the following time coordinate:
\begin{align}
    \tilde{t}_V \equiv t+\int \frac{a_7}{a_6}{\rm d}r 
    =t+\int \frac{\sqrt{1-A}}{A}\frac{\beta_V}{A+\beta_V}{\rm d}r~.
\end{align}
Using this time coordinate, the quadratic Lagrangian for the vector graviton can be written as
\begin{align}
     \frac{(j^2-2)(2l+1)}{2\pi j^2}\tilde{\mathcal{L}}_{2,V}=\tilde{a}_5(\partial_{\tilde{t}_V}a)^2-a_6(\partial_r a)^2 -a_8a^2~,\label{eq:veclag2}
\end{align}
where we have defined $\tilde{a}_5\equiv a_5+a_7^2/a_6$.
The tortoise coordinate for the vector graviton is defined by
\begin{align}
    \label{eq:vectorrs}r_{*,V}\equiv\int\frac{{\rm d}r}{F_V(r)}~, \quad
    F_V(r)\equiv\sqrt{\frac{a_6}{\tilde{a}_5}}
    =\frac{A+\beta_V}{\sqrt{1+\beta_V}}~.
\end{align}
Note that the horizon radius for the effective metric of the vector graviton, denoted by $r=r_V$ satisfies $F_V(r_V)=0$.
In terms of a new variable,
\begin{align}
    \Psi_V \equiv (\tilde{a}_5a_6)^{1/4}a~,
\end{align}
the master equation can be written in the form of a wave equation, i.e.,
\begin{align}
    \frac{\partial^2 \Psi_V}{\partial r_{*,V}^2}-\frac{\partial^2 \Psi_V}{\partial \tilde{t}_V^2}-V_{\mathrm{eff},V}(r)\Psi_V=0~,\label{eq:vecmas1}
\end{align}
where the effective potential for the vector graviton is given by
\begin{align}
    V_{\mathrm{eff},V}&=\frac{a_8}{\tilde{a}_5}+\frac{1}{2\sqrt{\tilde{a}_5a_6}}\frac{{\rm d}^2\sqrt{\tilde{a}_5a_6}}{{\rm d}r_{*,V}^2}-\frac{1}{4\tilde{a}_5a_6}\left(\frac{{\rm d}\sqrt{\tilde{a}_5a_6}}{{\rm d}r_{*,V}}\right)^2 \nonumber \\
    &=\sqrt{1+\beta_V}\,F_V\frac{j^2-2}{r^2}+\frac{2F_V}{r^2\gamma_1\sqrt{1+\beta_V}}\left\{\left[r(1-A)\gamma_1\right]'+r\gamma_2'\right\} \nonumber \\
    &\quad+\frac{F_V}{r^2\sqrt{\gamma_1}(1+\beta_V)^{1/4}}\left\{F_V\left[r^2\sqrt{\gamma_1}(1+\beta_V)^{1/4}\right]'\,\right\}'~. \label{eq:vecpot1}
\end{align}
In the frequency domain, the master equation reduces to
\begin{align}\label{eq:vectormasterfreq}
    \frac{{\rm d}^2Q_V}{{\rm d}r^2_{*,V}}
    +(\omega^2-V_{\mathrm{eff},V})
    Q_V=0~,
\end{align}
where we have denoted the master variable in the frequency domain as $Q_V$.
Similarly to the case of the tensor graviton, the frequency conjugate to $\tilde{t}_V$ coincides with that conjugate to $t$.

\subsection{Dipole perturbations}\label{sec:dipole}
In this subsection, we analyze the dynamics of dipole perturbations.
In particular, we argue that the dipole perturbations are ill-behaved unless $p_4=0$.

For $l=1$, the quadratic Lagrangian~(\ref{qlag1}) reduces to
    \begin{align}
    \frac{3}{4\pi}\mathcal{L}_2&=p_2h_1^2+p_3(\dot{h}_{1}-\partial_\rho h_0-p_4h_1+p_5\partial_\rho a)^2+p_6a^2+p_7ah_0 \nonumber \\
    &\quad+p_8h_1\partial_\rho a+p_9\dot{a}^2+p_{10}(\partial_\rho a)^2~.
    \end{align}
Varying this Lagrangian with respect to $h_0$, $h_1$, and $a$, we obtain the following set of equations of motion:
    \begin{align}
    2\partial_\rho \tilde{\chi}+p_7 a&=0~,\label{eq:dipeos1}\\
    2\dot{\tilde{\chi}}+2p_4\tilde{\chi}-p_8\partial_\rho a&=0~,\label{eq:dipeos2}\\
    \partial_\tau (p_9\partial_\tau a)+\partial_\rho (p_{10}\partial_\rho a)-p_6 a&=\frac{p_7}{2}h_0+\partial_\rho (p_5\tilde{\chi})~,\label{eq:dipeos3}
\end{align}
where we have defined $\tilde{\chi}\equiv p_3(-\partial_\rho h_0+p_5\partial_\rho a)$.
Note that we have imposed the gauge condition~$h_1=0$ after deriving the equations of motion.
[See the discussion in the paragraph below Eq.~\eqref{eq:vecpert}.]
Provided that $\partial_\rho p_4\ne 0$, one can remove $\tilde{\chi}$ and its derivatives from Eqs.~\eqref{eq:dipeos1} and \eqref{eq:dipeos2} to obtain
    \begin{align}
    \partial_\rho\left(\frac{\partial_\tau(p_7 a)+\partial_\rho(p_8\partial_\rho a)+p_4 p_7 a}{\partial_\rho p_4}\right)+p_7 a=0~,
    \end{align}
which contains at most first derivatives with respect to $\tau$ acting on $a$.
This implies that one cannot freely choose the initial condition for $\partial_\tau a$, and hence $a$ is not a propagating degree of freedom unless $\partial_\rho p_4=0$.
Meanwhile, provided that the metric function behaves as $A(r)=1+{\cal O}(1/r)$ and the EFT parameters~$\alpha$ and $M_3^2$ are constant at spatial infinity, which would be the case for an asymptotically Minkowski background, we find $p_4={\cal O}(r^{-3/2})$ unless $p_4$ vanishes identically.
[See Eq.~\eqref{ap:ps} for the expression of $p_4$.]
Therefore, in order to obtain a physically sensible dynamics of the dipole perturbations, one has to impose $p_4=0$.

Once the condition~$p_4=0$ is imposed on the dipole perturbations, we find that $p_7$ and $p_8$ are also vanishing, and hence Eqs.~\eqref{eq:dipeos1}--\eqref{eq:dipeos3} reduce to
    \begin{align}
    \partial_\rho\tilde{\chi}=0~, \quad
    \dot{\tilde{\chi}}=0~, \quad
    \partial_\tau (p_9\partial_\tau a)+\partial_\rho (p_{10}\partial_\rho a)-p_6 a=\partial_\rho (p_5\tilde{\chi})~.
    \end{align}
The first two equations imply $\tilde{\chi}={\rm const}$, and then the third equation describes the propagation of the vector graviton~$a$.
Note that the constant~$\tilde{\chi}$ corresponds to the slow rotation of a black hole.

\section{Quasinormal mode frequency}\label{sec:QNM}
In this section, we study QNMFs based on the master equations derived in the previous section.
As mentioned earlier in Subsection~\ref{ssec:odd}, we consider the EFT of vector-tensor gravity described by the action~\eqref{eq:oddeffectiveaction} with $\alpha=-M_3^2={\rm const}$ to ensure that the tensor and vector gravitons are decoupled for $l\ge 2$ and that we have dynamical dipole perturbations.
In this case, the tadpole cancellation condition~\eqref{alpha} implies that the metric function~$A(r)$ must be of the Schwarzschild(-de Sitter) form.
For this reason, in what follows, we restrict ourselves to the case of stealth Schwarzschild solution where
\begin{align}\label{eq:SchwarzschildAB}
    A(r)=B(r)=1-\frac{r_\mathrm{H}}{r}~,
\end{align}
with $r_{\rm H}$ corresponding to the horizon radius.

It should be noted that the QNM boundary conditions must be imposed at spatial infinity and the horizon for each degree of freedom.
Indeed, as mentioned in Subsections~\ref{sssec:tensor} and \ref{sssec:vector}, the horizon radii of the tensor/vector gravitons (denoted by $r_{T/V}$) are different from that of the background metric~$r_{\rm H}$ in general.
This happens when the propagation speeds of the tensor/vector gravitons are different from unity (i.e., $\alpha_T\ne 0$ for the tensor graviton and $\beta_V\ne 0$ for the vector graviton).
Therefore, the QNM boundary condition should be
    \begin{align}
    Q_T\sim
    \left\{\begin{array}{ll}
    e^{-i\omega r_{*,T}} & \quad (r\to r_T)~, \\
    e^{i\omega r_{*,T}} & \quad (r\to \infty)~,
    \end{array}\right.
    \end{align}
for the tensor graviton and
    \begin{align}
    Q_V\sim
    \left\{\begin{array}{ll}
    e^{-i\omega r_{*,V}} & \quad (r\to r_V)~, \\
    e^{i\omega r_{*,V}} & \quad (r\to \infty)~,
    \end{array}\right.
    \end{align}
for the vector graviton, where we use the tortoise coordinates~$r_{*, T/V}$ for each graviton defined in Eq.~\eqref{eq:tensorrs} and \eqref{eq:vectorrs} respectively.
Note that the effective potentials~$V_{{\rm eff},T/V}$~[given in Eq.~\eqref{eq:tenpot1} and \eqref{eq:vecpot1}] vanish at spatial infinity and $r=r_{T/V}$.

\subsection{Tensor graviton}
First, we consider the QNMFs for the tensor graviton based on Eq.~\eqref{eq:tensormastereqfreq}. 
For the Schwarzschild background~\eqref{eq:SchwarzschildAB}, the effective potential \eqref{eq:tenpot1} reads
\begin{align}\label{eq:SchwarzschildeffectiveVT}
V_{\mathrm{eff},T}(r) =
(1+\alpha_T)f_T(r)\left[\frac{l(l+1)}{r^2} - \frac{3r_T}{r^3} \right]~,
\end{align}
where
    \begin{align}
    f_T(r)\equiv (1+\alpha_T)^{-1/2}F_T(r)=1-\frac{r_T}{r}~, \quad
    r_T=\frac{r_{\rm H}}{1+\alpha_T}~.
    \end{align}
Also, the tortoise coordinate for the tensor graviton~\eqref{eq:tensorrs} can be written as
\begin{align}
    r_{*,T}=(1+\alpha_T)^{-1/2}\left[r+r_T \log\,\left| \frac{r}{r_T}-1\right|\,\right]~.
\end{align}
Recall that $\alpha_T$ is a constant given by Eq.~\eqref{eq:alpha} in terms of the EFT parameters.\footnote{Note that the LIGO/Virgo bound on the speed of gravitational waves implies that $|\alpha_T|\lesssim 10^{-15}$~\cite{LIGOScientific:2017vwq,LIGOScientific:2017ync,LIGOScientific:2017zic}.
Nevertheless, we will also consider $\alpha_T$ of ${\cal O}(10^{-1})$ for illustrative purposes.}
It should be noted that the conventional Regge-Wheeler potential is recovered when $\alpha_T=0$.
Note also that the effective potential~\eqref{eq:SchwarzschildeffectiveVT} is exactly the same as that of the stealth Schwarzschild solution in the EFT of scalar-tensor gravity~\cite{Mukohyama:2023xyf}, and hence the subsequent analysis proceeds in the same manner.
\begin{figure}
\centering
\includegraphics[width=.6\textwidth,clip]{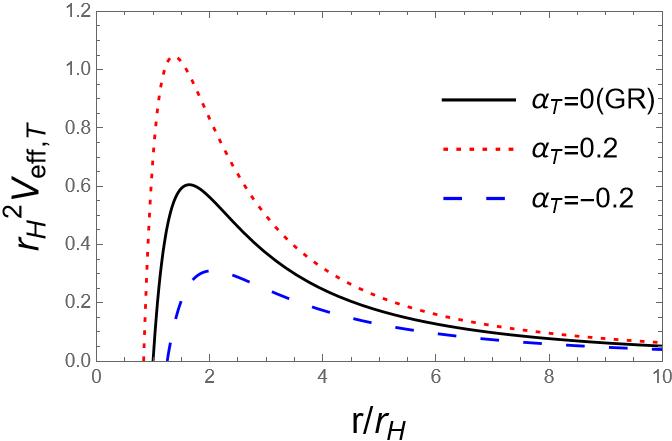}%
\caption{
Effective potential~$V_{{\rm eff},T}$ of the stealth Schwarzschild background for the $l=2$ tensor graviton as a function of $r/r_{\rm H}$.}
\label{fig:Schwarzschildgravitypoten}
\end{figure}
Figure~\ref{fig:Schwarzschildgravitypoten} shows the profile of the effective potential for different values of $\alpha_T$.
As expected, smaller values of $\alpha_T$ correspond to larger values of $r_T$, and vice versa.

Let us consider the following rescalings:
\begin{align}
    \tilde{\omega}_T&\equiv(1+\alpha_T)^{-1/2}\omega~,\\
    \tilde{r}_{*,T}&\equiv \sqrt{1+\alpha_T}\,r_{*,T}=r+r_T\log\,\left|\frac{r}{r_T}-1\right|~,\\
    \tilde{V}_{\mathrm{eff},T}&\equiv\frac{V_{\mathrm{eff},T}}{1+\alpha_T}=f_T(r)\left[\frac{l(l+1)}{r^2}-\frac{3r_T}{r^3}\right]~.
\end{align}
With these rescalings, it is straightforward to confirm that the master equation~\eqref{eq:tensormastereqfreq} reduces to the conventional Regge-Wheeler equation in GR, which implies
\begin{align}
    r_T\tilde{\omega}_T=r_\mathrm{H}\omega_\mathrm{GR}^{s=2}~.
\end{align}
Here, the right-hand side is the QNMF of odd-parity metric perturbations on the Schwarzschild background in GR normalized by $r_{\rm H}$, which is well known in the literature~\cite{Chandrasekhar:1975zza,Schutz:1985km,Mamani:2022akq}.
Going back to the original $\omega$, we obtain
\begin{align}
    \label{eq:relationT}r_\mathrm{H}\omega=r_\mathrm{H}\omega_\mathrm{GR}^{s=2}(1+\alpha_T)^{3/2}~.
\end{align}
Thus, for the Schwarzschild background, the QNMFs of the tensor graviton can be obtained from those in GR by simple scaling.
It should be noted that this relation holds for arbitrary multipoles and arbitrary overtones.

\subsection{Vector graviton}\label{sec:schvec}
Next, we consider the QNMFs for the vector graviton based on Eq.~\eqref{eq:vectormasterfreq}. 
For simplicity, we assume that the EFT parameters~$\gamma_1$ and $\gamma_2$ are constant, so that $\beta_V$ given by Eq.~\eqref{betaV} is also a constant.
For the Schwarzschild background~\eqref{eq:SchwarzschildAB}, the effective potential \eqref{eq:vecpot1} reads
\begin{align}\label{eq:SchwarzschildeffectiveV}
V_{\mathrm{eff},V}(r) = 
(1+\beta_V)f_V(r)\frac{l(l+1)}{r^2}~,
\end{align}
where
    \begin{align}
    f_V(r)\equiv (1+\beta_V)^{-1/2}F_V(r)=1-\frac{r_V}{r}~, \quad
    r_V=\frac{r_{\rm H}}{1+\beta_V}~.
    \end{align}
Also, the tortoise coordinate \eqref{eq:vectorrs} can be written as
\begin{align}
    r_{*,V}=(1+\beta_V)^{-1/2}\left[r+r_V \log\,\left| \frac{r}{r_V}-1\right|\,\right]~.
\end{align}
For $\beta_V=0$, the effective potential matches that of a spin-1 field in GR.
\begin{figure}
\centering
\includegraphics[width=.6\textwidth,clip]{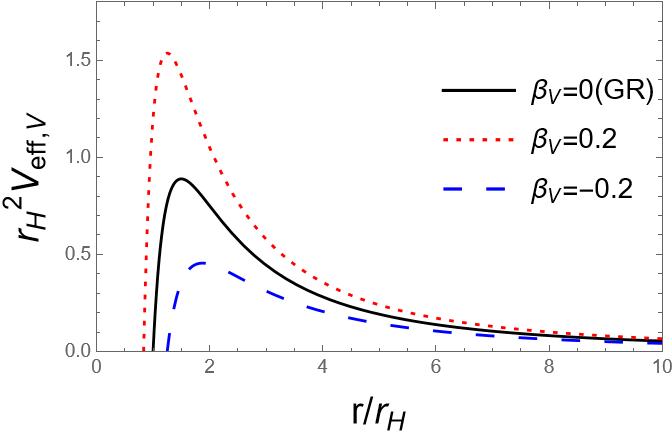}
\caption{
Effective potential~$V_{{\rm eff},V}$ of the stealth Schwarzschild background for the $l=2$ vector graviton as a function of $r/r_{\rm H}$.}
\label{fig:Schwarzschildvectorpoten}
\end{figure}
Figure~\ref{fig:Schwarzschildvectorpoten} shows the profile of the effective potential for different values of $\beta_V$.
As expected, smaller values of $\beta_V$ correspond to larger values of $r_V$, and vice versa.

Let us consider the following rescalings:
\begin{align}
    \tilde{\omega}_V&\equiv(1+\beta_V)^{-1/2}\omega~,\\
    \tilde{r}_{*,V}&\equiv \sqrt{1+\beta_V}\,r_{*,V}=r+r_V\log\,\left|\frac{r}{r_V}-1\right|~,\\
    \tilde{V}_{\mathrm{eff},V}&\equiv\frac{V_{\mathrm{eff},V}}{1+\beta_V}=f_V(r)\frac{l(l+1)}{r^2}~.
\end{align}
With these rescalings, it is straightforward to confirm that the master equation~\eqref{eq:vectormasterfreq} reduces to the conventional Regge-Wheeler equation in GR, which implies
\begin{align}
    r_V\tilde{\omega}_V=r_\mathrm{H}\omega_\mathrm{GR}^{s=1}~.
\end{align}
Here, the right-hand side is the QNMF of a spin-1 field on the Schwarzschild background in GR normalized by $r_{\rm H}$, which is well known in the literature~\cite{Mamani:2022akq}.
Going back to the original $\omega$, we obtain
\begin{align}
    \label{eq:relationV}r_\mathrm{H}\omega=r_\mathrm{H}\omega_\mathrm{GR}^{s=1}(1+\beta_V)^{3/2}~.
\end{align}
Similarly to the case of the tensor graviton, this relation holds for arbitrary multipoles and arbitrary overtones.

\begin{figure}
\centering
\includegraphics[width=.6\textwidth,clip]{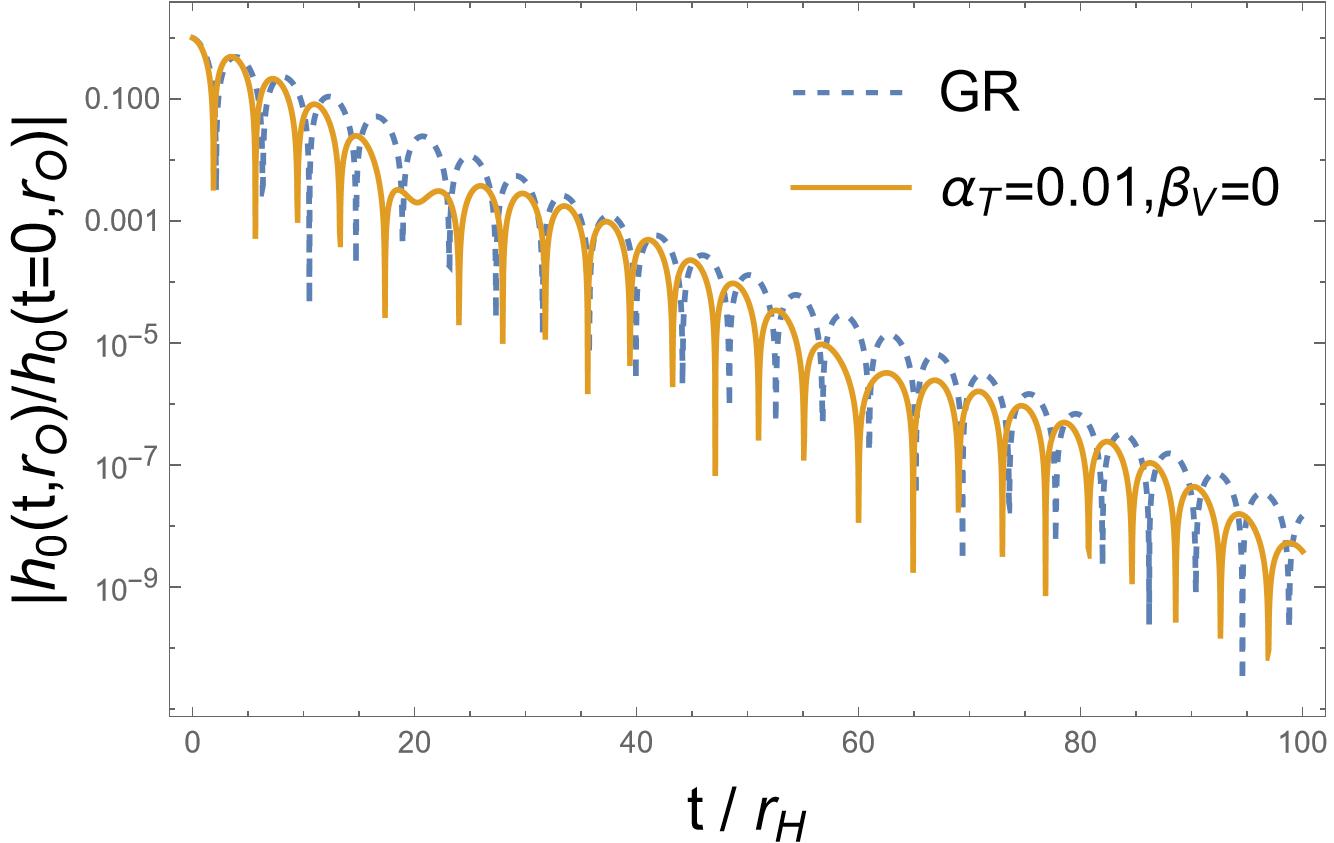}%
\caption{
Schematic behavior of $|h_0|$ measured at a distant region~$r=r_o$. The amplitude is normalized at $t=0$. We have chosen $\alpha_T=0.01$ and $\beta_V=0$, and taken into account only the $l=2$ fundamental QNMs of the tensor and vector gravitons. We see a non-trivial modulation in vector-tensor gravity (solid orange) due to a contribution of the vector graviton, which is in sharp contrast to the GR case (dashed blue).
}
\label{fig:QNM_beating_Schwarzschild}%
\end{figure}

\subsection{Modulation of ringdown gravitational waves}

Finally, let us comment on possible characteristic signals in observed gravitational waves.
As can be seen in Eq.~\eqref{h0h1}, the original metric perturbation variable~$h_0$ is a non-trivial linear combination of the two master variables~$\chi$ and $a$ in general.\footnote{As mentioned in Subsection~\ref{ssec:odd}, we have imposed the condition~$p_4=0$, under which $p_8$ is also vanishing, and therefore $h_1$ is not affected by the vector graviton~$a$.}
Since the tensor graviton~$\chi$ and the vector graviton~$a$ have different QNM spectra, a non-trivial modulation---such as a beating phenomenon---would be imprinted in the ringdown gravitational waves, as demonstrated in Figure~\ref{fig:QNM_beating_Schwarzschild}. 
This modulation appears in the time scale of $\sim 1/|\omega_{\chi} - \omega_{a}|$, where $\omega_{\chi}$ and $\omega_a$ are the real part of the fundamental QNMFs of the $\chi$ and $a$, respectively.

\section{Conclusions}\label{sec:conclusion}
We have studied odd-parity perturbations about static and spherically symmetric background based on the effective field theory (EFT) of vector-tensor gravity developed in \cite{Aoki:2023bmz}.
For perturbations with generic higher multipoles~$l\ge 2$, we have obtained the quadratic Lagrangian~\eqref{eq:lag4} in terms of the master variables~$\chi$ and $a$, which correspond to the tensor and vector gravitons, respectively.
Although $\chi$ and $a$ are coupled with each other in general, we have found that they can be decoupled on a stealth Schwarzschild(-de Sitter) background, and we have mainly focused on such a system for simplicity.
In this case, the dynamics of the tensor and vector gravitons are governed by Eq.~\eqref{eq:lag4'}, and we have derived the stability conditions as well as the master equation for each degree of freedom.

In Section \ref{sec:QNM}, we have computed quasinormal mode frequencies (QNMFs) 
of the stealth Schwarzschild background.
Similarly to the case of scalar-tensor gravity, we have found that the master equations can be reduced to those in GR upon a rescaling of parameters.
Therefore, the QNMFs can be obtained from those in GR by simple scaling [see Eqs.~\eqref{eq:relationT} and \eqref{eq:relationV}].
Since observed gravitational waves correspond to a linear combination of the two master variables, it is expected that a non-trivial modulation would show up in the ringdown gravitational waves, which is in sharp contrast to the case of scalar-tensor gravity.

It would be intriguing to investigate the observational distinguishability between scalar-tensor and vector-tensor EFTs.
Also, it is important to extend our analysis to the even-parity sector.
In the context of (gauged) ghost condensate~\cite{Arkani-Hamed:2003pdi,Cheng:2006us,Mukohyama:2006mm}, it has been known that a higher-derivative term, i.e., the scordatura term~\cite{Motohashi:2019ymr} is necessary and thus included from the beginning to render the perturbations weakly coupled, and this effect should be taken into account when we study even-parity black hole perturbations.\footnote{See Refs.~\cite{deRham:2019gha,Takahashi:2021bml} for the strong coupling problem in the context of black hole perturbations in scalar-tensor gravity without scordatura terms.
See also Ref.~\cite{Mukohyama:2025owu} for an analysis of spherical perturbations on an (approximately) stealth Schwarzschild background in the EFT of scalar-tensor gravity.}
Notice that the scordatura effect is implemented by default in our EFT of vector-tensor gravity, and therefore the EFT serves as an ideal framework to study stealth black hole solutions.
These issues are left for future work.

\acknowledgments
The work of H.K.~was supported by JST (Japan Science and Technology Agency) SPRING, Grant No.\ JPMJSP2110.
The work of N.O.~was supported by JSPS (Japan Society for the Promotion of Science) KAKENHI Grant No.\ JP23K13111 and by the Hakubi project at Kyoto University.
The work of K.T.~was supported in part by JSPS KAKENHI Grant No.\ JP23K13101.
The work of S.M.~was supported in part by JSPS KAKENHI Grant No.\ JP24K07017 and World Premier International Research Center Initiative (WPI), MEXT, Japan.

\section*{Appendix: Coefficients of the quadratic Lagrangian}\label{app}
\addcontentsline{toc}{section}{Appendix: Coefficients of the quadratic Lagrangian}
\renewcommand\theequation{A.\arabic{equation}}
\setcounter{equation}{0}

We summarize the coefficients of the quadratic Lagrangian of odd-parity perturbations introduced in Section~\ref{sec:perturbation}.
The coefficients~$p_i$ in the Lagrangian~\eqref{qlag1} are given as follows:
    \begin{align}
    &p_1=\frac{1}{2}(j^2-2)r^2\sqrt{1-A}(M_*^2+M_3^2)~, \quad
    p_2=-(j^2-2)\frac{M_*^2 r^2}{2\sqrt{1-A}}-p_3 p_4^2~, \nonumber \\
    &p_3=\frac{(M_*^2+M_3^2)r^4}{2\sqrt{1-A}}~, \quad
    p_4=\frac{(2-2A+rA')(\alpha+M_3^2)}{2r\sqrt{1-A}(M_*^2+M_3^2)}~, \quad
    p_5=-\frac{g_M M_3^2}{(1+g_M q)(M_*^2+M_3^2)}~, \nonumber \\
    &p_6=r^2\sqrt{1-A}\Bigg\{(j^2-2)\left[\frac{g_M^2 M_3^2}{2(1+g_M q)^2}-\gamma_2\right]+2\left[r(1-A)\gamma_1\right]'+2r\gamma_2' \nonumber \\
    &\qquad\qquad\qquad\quad -\frac{2g_M^2(1-A)(\alpha+M_3^2)}{(1+g_M q)^2}-\frac{g_M^2 M_*^2}{2(1+g_M q)^2}\left[r^4\left(\frac{1-A}{r^2}\right)'\,\right]'\Bigg\}~, \nonumber \\
    &p_7=\frac{g_M r^2\sqrt{1-A}}{1+g_M q}\left[(j^2-2)M_3^2-2(1-A)(M_3^2+\alpha)\right]~, \quad
    p_8=\frac{g_M M_*^2 r^4}{(1+g_M q)\sqrt{1-A}}\,p_4~, \nonumber \\
    &p_9=-r^4\sqrt{1-A}\,\gamma_1~, \quad
    p_{10}=-\frac{r^4\gamma_2}{\sqrt{1-A}}+\frac{g_M^2 M_*^2 M_3^2 r^4}{2(1+g_Mq)^2\sqrt{1-A}(M_*^2+M_3^2)}~. \label{ap:ps}
    \end{align}
The coefficients~$s_i$ in the Lagrangian~\eqref{eq:lag4} are as follows:
    \begin{align}
    \begin{split}
    &s_1=-\frac{(j^2-2)p_3^2}{p_2}~, \quad
    s_2=\frac{(j^2-2)p_3^2}{p_1}~, \\
    &s_3=(j^2-2)p_3\left[1+\frac{p_3p_4^2}{p_2}-\left(\frac{p_1+p_2}{p_1p_2}\dot{p}_3\right)^{\boldsymbol{\cdot}}-p_3\left(\frac{p_4}{p_2}\right)^{\boldsymbol{\cdot}}\,\right]~, \\
    &s_4=(j^2-2)p_9~, \quad
    s_5=(j^2-2)\left(\frac{p_8^2}{4p_2}-p_{10}\right)~, \quad
    s_6=(j^2-2)\left(\frac{p_7^2}{4p_1}-p_6\right)~, \\
    &s_7=\frac{(j^2-2)p_3p_8}{p_2}~, \quad
    s_8=(j^2-2)\left[p_3\left(2p_5+\frac{p_7}{p_1}\right)+\frac{p_8(\dot{p}_3+p_3p_4)}{p_2}\right]~, \\
    &s_9=-(j^2-2)p_3\left(\frac{p_7}{p_1}\right)^{\boldsymbol{\cdot}}~.
    \end{split}
    \end{align}

\bibliographystyle{utphys}
\bibliography{references}

@article{Aoki:2023bmz,
    author = "Aoki, Katsuki and Gorji, Mohammad Ali and Mukohyama, Shinji and Takahashi, Kazufumi and Yingcharoenrat, Vicharit",
    title = "{Effective field theory of black hole perturbations in vector-tensor gravity}",
    eprint = "2311.06767",
    archivePrefix = "arXiv",
    primaryClass = "hep-th",
    reportNumber = "YITP-23-138, IPMU23-0041",
    doi = "10.1088/1475-7516/2024/03/012",
    journal = "JCAP",
    volume = "03",
    pages = "012",
    year = "2024"
}

@article{Aoki:2021wew,
    author = "Aoki, Katsuki and Gorji, Mohammad Ali and Mukohyama, Shinji and Takahashi, Kazufumi",
    title = "{The effective field theory of vector-tensor theories}",
    eprint = "2111.08119",
    archivePrefix = "arXiv",
    primaryClass = "hep-th",
    reportNumber = "YITP-21-132, IPMU21-0079",
    doi = "10.1088/1475-7516/2022/01/059",
    journal = "JCAP",
    volume = "01",
    number = "01",
    pages = "059",
    year = "2022"
}

@article{Mukohyama:2022enj,
    author = "Mukohyama, Shinji and Yingcharoenrat, Vicharit",
    title = "{Effective field theory of black hole perturbations with timelike scalar profile: formulation}",
    eprint = "2204.00228",
    archivePrefix = "arXiv",
    primaryClass = "hep-th",
    reportNumber = "YITP-22-27, IPMU22-0012",
    doi = "10.1088/1475-7516/2022/09/010",
    journal = "JCAP",
    volume = "09",
    pages = "010",
    year = "2022"
}

@article{Mukohyama:2023xyf,
    author = "Mukohyama, Shinji and Takahashi, Kazufumi and Tomikawa, Keitaro and Yingcharoenrat, Vicharit",
    title = "{Quasinormal modes from EFT of black hole perturbations with timelike scalar profile}",
    eprint = "2304.14304",
    archivePrefix = "arXiv",
    primaryClass = "gr-qc",
    reportNumber = "YITP-23-45, IPMU23-0007, RUP-23-8",
    doi = "10.1088/1475-7516/2023/07/050",
    journal = "JCAP",
    volume = "07",
    pages = "050",
    year = "2023"
}

@article{Schutz:1985km,
    author = "Schutz, Bernard F. and Will, Clifford M.",
    title = "{BLACK HOLE NORMAL MODES: A SEMIANALYTIC APPROACH}",
    reportNumber = "PRINT-85-0063 (WASH.U.,ST.LOUIS)",
    doi = "10.1086/184453",
    journal = "Astrophys. J. Lett.",
    volume = "291",
    pages = "L33--L36",
    year = "1985"
}

@article{Mamani:2022akq,
    author = "Mamani, Luis A. H. and Masa, Angel D. D. and Sanches, Lucas Timotheo and Zanchin, Vilson T.",
    title = "{Revisiting the quasinormal modes of the Schwarzschild black hole: Numerical analysis}",
    eprint = "2206.03512",
    archivePrefix = "arXiv",
    primaryClass = "gr-qc",
    doi = "10.1140/epjc/s10052-022-10865-1",
    journal = "Eur. Phys. J. C",
    volume = "82",
    number = "10",
    pages = "897",
    year = "2022"
}

@article{Chandrasekhar:1975zza,
    author = "Chandrasekhar, S. and Detweiler, Steven L.",
    title = "{The quasi-normal modes of the Schwarzschild black hole}",
    doi = "10.1098/rspa.1975.0112",
    journal = "Proc. Roy. Soc. Lond. A",
    volume = "344",
    pages = "441--452",
    year = "1975"
}

@article{SupernovaSearchTeam:1998fmf,
    author = "Riess, Adam G. and others",
    collaboration = "Supernova Search Team",
    title = "{Observational evidence from supernovae for an accelerating universe and a cosmological constant}",
    eprint = "astro-ph/9805201",
    archivePrefix = "arXiv",
    doi = "10.1086/300499",
    journal = "Astron. J.",
    volume = "116",
    pages = "1009--1038",
    year = "1998"
}

@article{SupernovaCosmologyProject:1998vns,
    author = "Perlmutter, S. and others",
    collaboration = "Supernova Cosmology Project",
    title = "{Measurements of $\Omega$ and $\Lambda$ from 42 High Redshift Supernovae}",
    eprint = "astro-ph/9812133",
    archivePrefix = "arXiv",
    reportNumber = "LBNL-41801, LBL-41801",
    doi = "10.1086/307221",
    journal = "Astrophys. J.",
    volume = "517",
    pages = "565--586",
    year = "1999"
}

@article{Horndeski:1974wa,
    author = "Horndeski, Gregory Walter",
    title = "{Second-order scalar-tensor field equations in a four-dimensional space}",
    doi = "10.1007/BF01807638",
    journal = "Int. J. Theor. Phys.",
    volume = "10",
    pages = "363--384",
    year = "1974"
}

@article{Langlois:2015cwa,
    author = "Langlois, David and Noui, Karim",
    title = "{Degenerate higher derivative theories beyond Horndeski: evading the Ostrogradski instability}",
    eprint = "1510.06930",
    archivePrefix = "arXiv",
    primaryClass = "gr-qc",
    doi = "10.1088/1475-7516/2016/02/034",
    journal = "JCAP",
    volume = "02",
    pages = "034",
    year = "2016"
}

@article{Crisostomi:2016czh,
    author = "Crisostomi, Marco and Koyama, Kazuya and Tasinato, Gianmassimo",
    title = "{Extended Scalar-Tensor Theories of Gravity}",
    eprint = "1602.03119",
    archivePrefix = "arXiv",
    primaryClass = "hep-th",
    doi = "10.1088/1475-7516/2016/04/044",
    journal = "JCAP",
    volume = "04",
    pages = "044",
    year = "2016"
}

@article{BenAchour:2016fzp,
    author = "Ben Achour, Jibril and Crisostomi, Marco and Koyama, Kazuya and Langlois, David and Noui, Karim and Tasinato, Gianmassimo",
    title = "{Degenerate higher order scalar-tensor theories beyond Horndeski up to cubic order}",
    eprint = "1608.08135",
    archivePrefix = "arXiv",
    primaryClass = "hep-th",
    doi = "10.1007/JHEP12(2016)100",
    journal = "JHEP",
    volume = "12",
    pages = "100",
    year = "2016"
}

@article{Takahashi:2017pje,
    author = "Takahashi, Kazufumi and Kobayashi, Tsutomu",
    title = "{Extended mimetic gravity: Hamiltonian analysis and gradient instabilities}",
    eprint = "1708.02951",
    archivePrefix = "arXiv",
    primaryClass = "gr-qc",
    reportNumber = "RESCEU-9-17, RUP-17-15",
    doi = "10.1088/1475-7516/2017/11/038",
    journal = "JCAP",
    volume = "11",
    pages = "038",
    year = "2017"
}

@article{Langlois:2018jdg,
    author = "Langlois, David and Mancarella, Michele and Noui, Karim and Vernizzi, Filippo",
    title = "{Mimetic gravity as DHOST theories}",
    eprint = "1802.03394",
    archivePrefix = "arXiv",
    primaryClass = "gr-qc",
    doi = "10.1088/1475-7516/2019/02/036",
    journal = "JCAP",
    volume = "02",
    pages = "036",
    year = "2019"
}

@article{DeFelice:2018ewo,
    author = "De Felice, Antonio and Langlois, David and Mukohyama, Shinji and Noui, Karim and Wang, Anzhong",
    title = "{Generalized instantaneous modes in higher-order scalar-tensor theories}",
    eprint = "1803.06241",
    archivePrefix = "arXiv",
    primaryClass = "hep-th",
    doi = "10.1103/PhysRevD.98.084024",
    journal = "Phys. Rev. D",
    volume = "98",
    number = "8",
    pages = "084024",
    year = "2018"
}

@article{DeFelice:2021hps,
    author = "De Felice, Antonio and Mukohyama, Shinji and Takahashi, Kazufumi",
    title = "{Nonlinear definition of the shadowy mode in higher-order scalar-tensor theories}",
    eprint = "2110.03194",
    archivePrefix = "arXiv",
    primaryClass = "gr-qc",
    reportNumber = "YITP-21-104, IPMU21-0061",
    doi = "10.1088/1475-7516/2021/12/020",
    journal = "JCAP",
    volume = "12",
    number = "12",
    pages = "020",
    year = "2021"
}

@article{Takahashi:2021ttd,
    author = "Takahashi, Kazufumi and Motohashi, Hayato and Minamitsuji, Masato",
    title = "{Invertible disformal transformations with higher derivatives}",
    eprint = "2111.11634",
    archivePrefix = "arXiv",
    primaryClass = "gr-qc",
    reportNumber = "YITP-21-137",
    doi = "10.1103/PhysRevD.105.024015",
    journal = "Phys. Rev. D",
    volume = "105",
    number = "2",
    pages = "024015",
    year = "2022"
}

@article{Takahashi:2022mew,
    author = "Takahashi, Kazufumi and Minamitsuji, Masato and Motohashi, Hayato",
    title = "{Generalized disformal Horndeski theories: Cosmological perturbations and consistent matter coupling}",
    eprint = "2209.02176",
    archivePrefix = "arXiv",
    primaryClass = "gr-qc",
    reportNumber = "YITP-22-89",
    doi = "10.1093/ptep/ptac161",
    journal = "PTEP",
    volume = "2023",
    number = "1",
    pages = "013E01",
    year = "2023"
}

@article{Takahashi:2023jro,
    author = "Takahashi, Kazufumi and Minamitsuji, Masato and Motohashi, Hayato",
    title = "{Effective description of generalized disformal theories}",
    eprint = "2304.08624",
    archivePrefix = "arXiv",
    primaryClass = "gr-qc",
    reportNumber = "YITP-23-46",
    doi = "10.1088/1475-7516/2023/07/009",
    journal = "JCAP",
    volume = "07",
    pages = "009",
    year = "2023"
}

@article{Heisenberg:2014rta,
    author = "Heisenberg, Lavinia",
    title = "{Generalization of the Proca Action}",
    eprint = "1402.7026",
    archivePrefix = "arXiv",
    primaryClass = "hep-th",
    doi = "10.1088/1475-7516/2014/05/015",
    journal = "JCAP",
    volume = "05",
    pages = "015",
    year = "2014"
}

@article{Tasinato:2014eka,
    author = "Tasinato, Gianmassimo",
    title = "{Cosmic Acceleration from Abelian Symmetry Breaking}",
    eprint = "1402.6450",
    archivePrefix = "arXiv",
    primaryClass = "hep-th",
    doi = "10.1007/JHEP04(2014)067",
    journal = "JHEP",
    volume = "04",
    pages = "067",
    year = "2014"
}

@article{Allys:2015sht,
    author = "Allys, Erwan and Peter, Patrick and Rodriguez, Yeinzon",
    title = "{Generalized Proca action for an Abelian vector field}",
    eprint = "1511.03101",
    archivePrefix = "arXiv",
    primaryClass = "hep-th",
    reportNumber = "PI-UAN-2015-589FT",
    doi = "10.1088/1475-7516/2016/02/004",
    journal = "JCAP",
    volume = "02",
    pages = "004",
    year = "2016"
}

@article{BeltranJimenez:2016rff,
    author = "Beltran Jimenez, Jose and Heisenberg, Lavinia",
    title = "{Derivative self-interactions for a massive vector field}",
    eprint = "1602.03410",
    archivePrefix = "arXiv",
    primaryClass = "hep-th",
    doi = "10.1016/j.physletb.2016.04.017",
    journal = "Phys. Lett. B",
    volume = "757",
    pages = "405--411",
    year = "2016"
}

@article{Allys:2016jaq,
    author = "Allys, Erwan and Beltran Almeida, Juan P. and Peter, Patrick and Rodr\'\i{}guez, Yeinzon",
    title = "{On the 4D generalized Proca action for an Abelian vector field}",
    eprint = "1605.08355",
    archivePrefix = "arXiv",
    primaryClass = "hep-th",
    reportNumber = "PI-UAN-2016-595FT",
    doi = "10.1088/1475-7516/2016/09/026",
    journal = "JCAP",
    volume = "09",
    pages = "026",
    year = "2016"
}

@article{Heisenberg:2016eld,
    author = "Heisenberg, Lavinia and Kase, Ryotaro and Tsujikawa, Shinji",
    title = "{Beyond generalized Proca theories}",
    eprint = "1605.05565",
    archivePrefix = "arXiv",
    primaryClass = "hep-th",
    doi = "10.1016/j.physletb.2016.07.052",
    journal = "Phys. Lett. B",
    volume = "760",
    pages = "617--626",
    year = "2016"
}

@article{deRham:2020yet,
    author = "de Rham, Claudia and Pozsgay, Victor",
    title = "{New class of Proca interactions}",
    eprint = "2003.13773",
    archivePrefix = "arXiv",
    primaryClass = "hep-th",
    reportNumber = "Imperial/TP/2020/CdR/01",
    doi = "10.1103/PhysRevD.102.083508",
    journal = "Phys. Rev. D",
    volume = "102",
    number = "8",
    pages = "083508",
    year = "2020"
}

@article{Kimura:2016rzw,
    author = "Kimura, Rampei and Naruko, Atsushi and Yoshida, Daisuke",
    title = "{Extended vector-tensor theories}",
    eprint = "1608.07066",
    archivePrefix = "arXiv",
    primaryClass = "gr-qc",
    doi = "10.1088/1475-7516/2017/01/002",
    journal = "JCAP",
    volume = "01",
    pages = "002",
    year = "2017"
}

@article{LIGOScientific:2016aoc,
    author = "Abbott, B. P. and others",
    collaboration = "LIGO Scientific, Virgo",
    title = "{Observation of Gravitational Waves from a Binary Black Hole Merger}",
    eprint = "1602.03837",
    archivePrefix = "arXiv",
    primaryClass = "gr-qc",
    reportNumber = "LIGO-P150914",
    doi = "10.1103/PhysRevLett.116.061102",
    journal = "Phys. Rev. Lett.",
    volume = "116",
    number = "6",
    pages = "061102",
    year = "2016"
}

@article{LISA:2024hlh,
    author = "Colpi, Monica and others",
    collaboration = "LISA",
    title = "{LISA Definition Study Report}",
    eprint = "2402.07571",
    archivePrefix = "arXiv",
    primaryClass = "astro-ph.CO",
    month = "2",
    year = "2024"
}

@article{Arkani-Hamed:2003pdi,
    author = "Arkani-Hamed, Nima and Cheng, Hsin-Chia and Luty, Markus A. and Mukohyama, Shinji",
    title = "{Ghost condensation and a consistent infrared modification of gravity}",
    eprint = "hep-th/0312099",
    archivePrefix = "arXiv",
    reportNumber = "HUTP-03-A081, UMD-PPP-04-012",
    doi = "10.1088/1126-6708/2004/05/074",
    journal = "JHEP",
    volume = "05",
    pages = "074",
    year = "2004"
}

@article{Arkani-Hamed:2003juy,
    author = "Arkani-Hamed, Nima and Creminelli, Paolo and Mukohyama, Shinji and Zaldarriaga, Matias",
    title = "{Ghost inflation}",
    eprint = "hep-th/0312100",
    archivePrefix = "arXiv",
    reportNumber = "HUTP-03-A079",
    doi = "10.1088/1475-7516/2004/04/001",
    journal = "JCAP",
    volume = "04",
    pages = "001",
    year = "2004"
}

@article{Cheung:2007st,
    author = "Cheung, Clifford and Creminelli, Paolo and Fitzpatrick, A. Liam and Kaplan, Jared and Senatore, Leonardo",
    title = "{The Effective Field Theory of Inflation}",
    eprint = "0709.0293",
    archivePrefix = "arXiv",
    primaryClass = "hep-th",
    reportNumber = "IC-2007-032",
    doi = "10.1088/1126-6708/2008/03/014",
    journal = "JHEP",
    volume = "03",
    pages = "014",
    year = "2008"
}

@article{Gubitosi:2012hu,
    author = "Gubitosi, Giulia and Piazza, Federico and Vernizzi, Filippo",
    title = "{The Effective Field Theory of Dark Energy}",
    eprint = "1210.0201",
    archivePrefix = "arXiv",
    primaryClass = "hep-th",
    doi = "10.1088/1475-7516/2013/02/032",
    journal = "JCAP",
    volume = "02",
    pages = "032",
    year = "2013"
}

@article{Barura:2024uog,
    author = "Barura, Chams Gharib Ali and Kobayashi, Hajime and Mukohyama, Shinji and Oshita, Naritaka and Takahashi, Kazufumi and Yingcharoenrat, Vicharit",
    title = "{Tidal Love numbers from EFT of black hole perturbations with timelike scalar profile}",
    eprint = "2405.10813",
    archivePrefix = "arXiv",
    primaryClass = "gr-qc",
    reportNumber = "YITP-24-62, IPMU24-0018, RIKEN-iTHEMS-Report-24",
    doi = "10.1088/1475-7516/2024/09/001",
    journal = "JCAP",
    volume = "09",
    pages = "001",
    year = "2024"
}

@article{LIGOScientific:2017vwq,
    author = "Abbott, B. P. and others",
    collaboration = "LIGO Scientific, Virgo",
    title = "{GW170817: Observation of Gravitational Waves from a Binary Neutron Star Inspiral}",
    eprint = "1710.05832",
    archivePrefix = "arXiv",
    primaryClass = "gr-qc",
    reportNumber = "LIGO-P170817",
    doi = "10.1103/PhysRevLett.119.161101",
    journal = "Phys. Rev. Lett.",
    volume = "119",
    number = "16",
    pages = "161101",
    year = "2017"
}

@article{LIGOScientific:2017ync,
    author = "Abbott, B. P. and others",
    collaboration = "LIGO Scientific, Virgo, Fermi GBM, INTEGRAL, IceCube, AstroSat Cadmium Zinc Telluride Imager Team, IPN, Insight-Hxmt, ANTARES, Swift, AGILE Team, 1M2H Team, Dark Energy Camera GW-EM, DES, DLT40, GRAWITA, Fermi-LAT, ATCA, ASKAP, Las Cumbres Observatory Group, OzGrav, DWF (Deeper Wider Faster Program), AST3, CAASTRO, VINROUGE, MASTER, J-GEM, GROWTH, JAGWAR, CaltechNRAO, TTU-NRAO, NuSTAR, Pan-STARRS, MAXI Team, TZAC Consortium, KU, Nordic Optical Telescope, ePESSTO, GROND, Texas Tech University, SALT Group, TOROS, BOOTES, MWA, CALET, IKI-GW Follow-up, H.E.S.S., LOFAR, LWA, HAWC, Pierre Auger, ALMA, Euro VLBI Team, Pi of Sky, Chandra Team at McGill University, DFN, ATLAS Telescopes, High Time Resolution Universe Survey, RIMAS, RATIR, SKA South Africa/MeerKAT",
    title = "{Multi-messenger Observations of a Binary Neutron Star Merger}",
    eprint = "1710.05833",
    archivePrefix = "arXiv",
    primaryClass = "astro-ph.HE",
    reportNumber = "LIGO-P1700294, VIR-0802A-17, FERMILAB-PUB-17-478-A-AE-CD",
    doi = "10.3847/2041-8213/aa91c9",
    journal = "Astrophys. J. Lett.",
    volume = "848",
    number = "2",
    pages = "L12",
    year = "2017"
}

@article{LIGOScientific:2017zic,
    author = "Abbott, B. P. and others",
    collaboration = "LIGO Scientific, Virgo, Fermi-GBM, INTEGRAL",
    title = "{Gravitational Waves and Gamma-rays from a Binary Neutron Star Merger: GW170817 and GRB 170817A}",
    eprint = "1710.05834",
    archivePrefix = "arXiv",
    primaryClass = "astro-ph.HE",
    reportNumber = "LIGO-P1700308",
    doi = "10.3847/2041-8213/aa920c",
    journal = "Astrophys. J. Lett.",
    volume = "848",
    number = "2",
    pages = "L13",
    year = "2017"
}

@article{Mukohyama:2022skk,
    author = "Mukohyama, Shinji and Takahashi, Kazufumi and Yingcharoenrat, Vicharit",
    title = "{Generalized Regge-Wheeler equation from Effective Field Theory of black hole perturbations with a timelike scalar profile}",
    eprint = "2208.02943",
    archivePrefix = "arXiv",
    primaryClass = "gr-qc",
    reportNumber = "YITP-22-78, IPMU22-0039",
    doi = "10.1088/1475-7516/2022/10/050",
    journal = "JCAP",
    volume = "10",
    pages = "050",
    year = "2022"
}

@article{Regge:1957td,
    author = "Regge, Tullio and Wheeler, John A.",
    title = "{Stability of a Schwarzschild singularity}",
    doi = "10.1103/PhysRev.108.1063",
    journal = "Phys. Rev.",
    volume = "108",
    pages = "1063--1069",
    year = "1957"
}

@article{DeFelice:2022xvq,
    author = "De Felice, Antonio and Mukohyama, Shinji and Takahashi, Kazufumi",
    title = "{Avoidance of Strong Coupling in General Relativity Solutions with a Timelike Scalar Profile in a Class of Ghost-Free Scalar-Tensor Theories}",
    eprint = "2204.02032",
    archivePrefix = "arXiv",
    primaryClass = "gr-qc",
    reportNumber = "YITP-22-25, IPMU22-0014",
    doi = "10.1103/PhysRevLett.129.031103",
    journal = "Phys. Rev. Lett.",
    volume = "129",
    number = "3",
    pages = "031103",
    year = "2022"
}

@article{Takahashi:2023vva,
    author = "Takahashi, Kazufumi",
    title = "{Invertible disformal transformations with arbitrary higher-order derivatives}",
    eprint = "2307.08814",
    archivePrefix = "arXiv",
    primaryClass = "gr-qc",
    reportNumber = "YITP-23-91",
    doi = "10.1103/PhysRevD.108.084031",
    journal = "Phys. Rev. D",
    volume = "108",
    number = "8",
    pages = "084031",
    year = "2023"
}

@article{Cheng:2006us,
    author = "Cheng, Hsin-Chia and Luty, Markus A. and Mukohyama, Shinji and Thaler, Jesse",
    title = "{Spontaneous Lorentz breaking at high energies}",
    eprint = "hep-th/0603010",
    archivePrefix = "arXiv",
    reportNumber = "HUTP-06-A0006, UTAP-551",
    doi = "10.1088/1126-6708/2006/05/076",
    journal = "JHEP",
    volume = "05",
    pages = "076",
    year = "2006"
}

@article{Chagoya:2016aar,
    author = "Chagoya, Javier and Niz, Gustavo and Tasinato, Gianmassimo",
    title = "{Black Holes and Abelian Symmetry Breaking}",
    eprint = "1602.08697",
    archivePrefix = "arXiv",
    primaryClass = "hep-th",
    doi = "10.1088/0264-9381/33/17/175007",
    journal = "Class. Quant. Grav.",
    volume = "33",
    number = "17",
    pages = "175007",
    year = "2016"
}

@article{Minamitsuji:2017aan,
    author = "Minamitsuji, Masato",
    title = "{Black holes in the generalized Proca theory}",
    doi = "10.1007/s10714-017-2250-7",
    journal = "Gen. Rel. Grav.",
    volume = "49",
    number = "7",
    pages = "86",
    year = "2017"
}

@article{Minamitsuji:2021gcq,
    author = "Minamitsuji, Masato",
    title = "{Black holes in the quadratic-order extended vector-tensor theories}",
    eprint = "2105.08936",
    archivePrefix = "arXiv",
    primaryClass = "gr-qc",
    doi = "10.1088/1361-6382/abed62",
    journal = "Class. Quant. Grav.",
    volume = "38",
    number = "10",
    pages = "105011",
    year = "2021"
}

@article{Mukohyama:2005rw,
    author = "Mukohyama, Shinji",
    title = "{Black holes in the ghost condensate}",
    eprint = "hep-th/0502189",
    archivePrefix = "arXiv",
    reportNumber = "UTAP-512, RESCEU-2-05, HUTP-05-A0009",
    doi = "10.1103/PhysRevD.71.104019",
    journal = "Phys. Rev. D",
    volume = "71",
    pages = "104019",
    year = "2005"
}

@article{Babichev:2013cya,
    author = "Babichev, Eugeny and Charmousis, Christos",
    title = "{Dressing a black hole with a time-dependent Galileon}",
    eprint = "1312.3204",
    archivePrefix = "arXiv",
    primaryClass = "gr-qc",
    reportNumber = "LPT-ORSAY-13-105",
    doi = "10.1007/JHEP08(2014)106",
    journal = "JHEP",
    volume = "08",
    pages = "106",
    year = "2014"
}

@article{Kobayashi:2014eva,
    author = "Kobayashi, Tsutomu and Tanahashi, Norihiro",
    title = "{Exact black hole solutions in shift symmetric scalar\textendash{}tensor theories}",
    eprint = "1403.4364",
    archivePrefix = "arXiv",
    primaryClass = "gr-qc",
    reportNumber = "RUP-14-7, IPMU14-0057",
    doi = "10.1093/ptep/ptu096",
    journal = "PTEP",
    volume = "2014",
    pages = "073E02",
    year = "2014"
}

@article{BenAchour:2018dap,
    author = "Ben Achour, Jibril and Liu, Hongguang",
    title = "{Hairy Schwarzschild-(A)dS black hole solutions in degenerate higher order scalar-tensor theories beyond shift symmetry}",
    eprint = "1811.05369",
    archivePrefix = "arXiv",
    primaryClass = "gr-qc",
    doi = "10.1103/PhysRevD.99.064042",
    journal = "Phys. Rev. D",
    volume = "99",
    number = "6",
    pages = "064042",
    year = "2019"
}

@article{Motohashi:2018wdq,
    author = "Motohashi, Hayato and Minamitsuji, Masato",
    title = "{General Relativity solutions in modified gravity}",
    eprint = "1804.01731",
    archivePrefix = "arXiv",
    primaryClass = "gr-qc",
    reportNumber = "YITP-18-27",
    doi = "10.1016/j.physletb.2018.04.041",
    journal = "Phys. Lett. B",
    volume = "781",
    pages = "728--734",
    year = "2018"
}

@article{Motohashi:2019sen,
    author = "Motohashi, Hayato and Minamitsuji, Masato",
    title = "{Exact black hole solutions in shift-symmetric quadratic degenerate higher-order scalar-tensor theories}",
    eprint = "1901.04658",
    archivePrefix = "arXiv",
    primaryClass = "gr-qc",
    reportNumber = "YITP-19-01",
    doi = "10.1103/PhysRevD.99.064040",
    journal = "Phys. Rev. D",
    volume = "99",
    number = "6",
    pages = "064040",
    year = "2019"
}

@article{Takahashi:2020hso,
    author = "Takahashi, Kazufumi and Motohashi, Hayato",
    title = "{General Relativity solutions with stealth scalar hair in quadratic higher-order scalar-tensor theories}",
    eprint = "2004.03883",
    archivePrefix = "arXiv",
    primaryClass = "gr-qc",
    reportNumber = "KOBE-COSMO-20-05, YITP-20-35",
    doi = "10.1088/1475-7516/2020/06/034",
    journal = "JCAP",
    volume = "06",
    pages = "034",
    year = "2020"
}

@article{Motohashi:2016prk,
    author = "Motohashi, Hayato and Suyama, Teruaki and Takahashi, Kazufumi",
    title = "{Fundamental theorem on gauge fixing at the action level}",
    eprint = "1608.00071",
    archivePrefix = "arXiv",
    primaryClass = "gr-qc",
    reportNumber = "RESCEU-26-16",
    doi = "10.1103/PhysRevD.94.124021",
    journal = "Phys. Rev. D",
    volume = "94",
    number = "12",
    pages = "124021",
    year = "2016"
}

@article{Aoki:2024ktc,
    author = "Aoki, Katsuki and Gorji, Mohammad Ali and Hiramatsu, Takashi and Mukohyama, Shinji and Pookkillath, Masroor C. and Takahashi, Kazufumi",
    title = "{CMB spectrum in unified EFT of dark energy: scalar-tensor and vector-tensor theories}",
    eprint = "2405.04265",
    archivePrefix = "arXiv",
    primaryClass = "astro-ph.CO",
    reportNumber = "YITP-24-56, RUP-24-8, IPMU24-0016",
    doi = "10.1088/1475-7516/2024/07/056",
    journal = "JCAP",
    volume = "07",
    pages = "056",
    year = "2024"
}

@article{Cardoso:2024qie,
    author = "Cardoso, Vitor and Mukohyama, Shinji and Oshita, Naritaka and Takahashi, Kazufumi",
    title = "{Black holes, multiple propagation speeds, and energy extraction}",
    eprint = "2404.05790",
    archivePrefix = "arXiv",
    primaryClass = "gr-qc",
    reportNumber = "YITP-24-44, IPMU24-0014, RIKEN-iTHEMS-Report-24",
    doi = "10.1103/PhysRevD.109.124036",
    journal = "Phys. Rev. D",
    volume = "109",
    number = "12",
    pages = "124036",
    year = "2024"
}

@article{Mukohyama:2006mm,
    author = "Mukohyama, Shinji",
    title = "{Towards a Higgs phase of gravity in string theory}",
    eprint = "hep-th/0610254",
    archivePrefix = "arXiv",
    reportNumber = "UTAP-569, RESCEU-26-06",
    doi = "10.1088/1126-6708/2007/05/048",
    journal = "JHEP",
    volume = "05",
    pages = "048",
    year = "2007"
}

@article{Mukohyama:2025owu,
    author = "Mukohyama, Shinji and Takahashi, Kazufumi and Tomikawa, Keitaro and Yingcharoenrat, Vicharit",
    title = "{Spherical black hole perturbations in EFT of scalar-tensor gravity with timelike scalar profile}",
    eprint = "2503.00520",
    archivePrefix = "arXiv",
    primaryClass = "gr-qc",
    reportNumber = "YITP-25-30, IPMU25-0009",
    doi = "10.1088/1475-7516/2025/05/084",
    journal = "JCAP",
    volume = "05",
    pages = "084",
    year = "2025"
}

@article{deRham:2019gha,
    author = "de Rham, Claudia and Zhang, Jun",
    title = "{Perturbations of stealth black holes in degenerate higher-order scalar-tensor theories}",
    eprint = "1907.00699",
    archivePrefix = "arXiv",
    primaryClass = "hep-th",
    reportNumber = "Imperial/TP/2019/CdR/03",
    doi = "10.1103/PhysRevD.100.124023",
    journal = "Phys. Rev. D",
    volume = "100",
    number = "12",
    pages = "124023",
    year = "2019"
}

@article{Takahashi:2021bml,
    author = "Takahashi, Kazufumi and Motohashi, Hayato",
    title = "{Black hole perturbations in DHOST theories: master variables, gradient instability, and strong coupling}",
    eprint = "2106.07128",
    archivePrefix = "arXiv",
    primaryClass = "gr-qc",
    reportNumber = "YITP-21-38",
    doi = "10.1088/1475-7516/2021/08/013",
    journal = "JCAP",
    volume = "08",
    pages = "013",
    year = "2021"
}

@article{Motohashi:2019ymr,
    author = "Motohashi, Hayato and Mukohyama, Shinji",
    title = "{Weakly-coupled stealth solution in scordatura degenerate theory}",
    eprint = "1912.00378",
    archivePrefix = "arXiv",
    primaryClass = "gr-qc",
    reportNumber = "YITP-19-98, IPMU19-0156",
    doi = "10.1088/1475-7516/2020/01/030",
    journal = "JCAP",
    volume = "01",
    pages = "030",
    year = "2020"
}

@article{Atkins:2023axs,
    author = "Atkins, Bill and Tasinato, Gianmassimo",
    title = "{Hidden conformal symmetries for black holes in modified gravity}",
    eprint = "2311.03860",
    archivePrefix = "arXiv",
    primaryClass = "gr-qc",
    doi = "10.1103/PhysRevD.108.104070",
    journal = "Phys. Rev. D",
    volume = "108",
    number = "10",
    pages = "104070",
    year = "2023"
}

\end{document}